\documentclass[11pt,a4paper]{article}

\usepackage{ascmac}
\usepackage{amsmath}
\usepackage{amssymb}
\usepackage{bm}
\usepackage[footnotesize,bf]{caption}
\usepackage{fullpage}
\usepackage{color}
\usepackage{float}
\usepackage{graphicx}
\usepackage[]{natbib}
\usepackage{subfigure}
\usepackage{txfonts}
\usepackage{threeparttable}
\usepackage{wrapfig}
\usepackage[]{multicol}
\usepackage{wrapfig}
\usepackage{authblk}

\usepackage{floatflt}

\title{\large Complete list of the ASTRO-H Science Working Group}
\date{\vspace{-0.5cm}}
\newcommand{\MakeWhitePaperTitle}{
	\begin{center}
		\begin{figure}
			\vspace{1cm}
			\begin{center}
				\includegraphics[width=0.2\hsize]{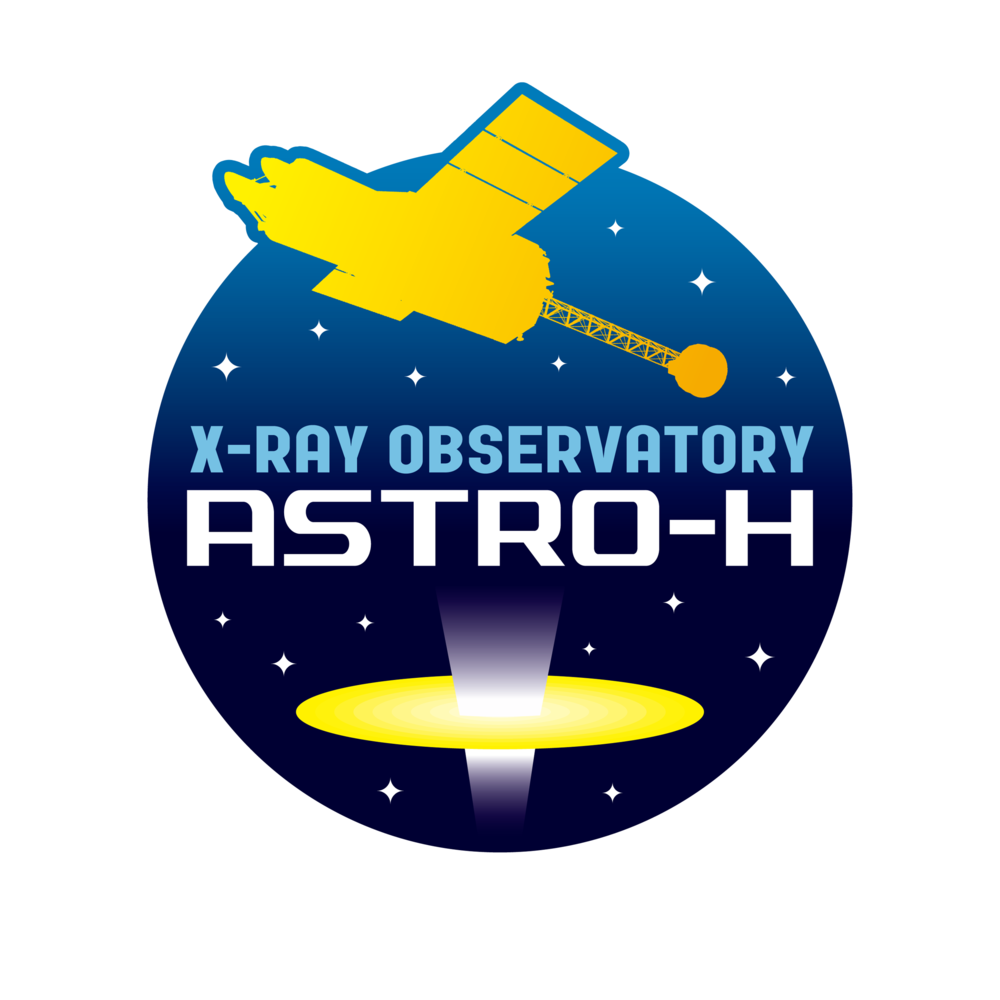}
			\end{center}
		\end{figure}
		\vspace{1cm}
		{\LARGE
		ASTRO-H Space X-ray Observatory\\
		White Paper\\
		}
		\vspace{5mm}
		{\large
		\WhitePaperTitle\\
		}
		\vspace{1cm}
		{
		\WhitePaperAuthors\\
		on behalf of the ASTRO-H Science Working Group
		}
	\end{center}
}

\usepackage{authblk}
\author[a]{Tadayuki~Takahashi}
\author[a]{Kazuhisa~Mitsuda}
\author[b]{Richard~Kelley}
\author[c]{Felix~Aharonian}
\author[d]{Hiroki~Akamatsu}
\author[e]{Fumie~Akimoto}
\author[f]{Steve~Allen}
\author[g]{Naohisa~Anabuki}
\author[b]{Lorella~Angelini}
\author[h]{Keith~Arnaud}
% ...
\affil[a]{Institute of Space and Astronautical Science (ISAS), Japan Aerospace Exploration Agency (JAXA), Kanagawa 252-5210, Japan}
\affil[b]{NASA/Goddard Space Flight Center, MD 20771, USA}
\affil[c]{Astronomy and Astrophysics Section, Dublin Institute for Advanced Studies, Dublin 2, Ireland}
\affil[d]{SRON Netherlands Institute for Space Research, Utrecht, The Netherlands}
\affil[e]{Department of Physics, Nagoya University, Aichi 338-8570, Japan}
\affil[f]{Kavli Institute for Particle Astrophysics and Cosmology, Stanford University, CA 94305, USA}
\affil[g]{Department of Earth and Space Science, Osaka University, Osaka 560-0043, Japan}
\affil[h]{Department of Astronomy, University of Maryland, MD 20742, USA}
% ...

\begin{document}

\newcommand{\WhitePaperTitle}{AGN Winds}
\newcommand{\WhitePaperAuthors}{
	J.~S.~Kaastra~(SRON),
	Y.~Terashima~(Ehime~University),
	T.~Kallman~(NASA/GSFC),
	Y.~Haba~(Aichi~University~of~Education),
	E.~Costantini~(SRON),
	L.~Gallo~(Saint~Mary's~University),
	Y.~Fukazawa~(Hiroshima~University),
	F.~Tombesi~(University~of~Maryland),
	N.~Anabuki~(Osaka~University),
	H.~Awaki~(Ehime~University),
	G.~Brown~(LLNL),
	L.~di~Gesu~(SRON),
	K.~Ebisawa~(JAXA),
	J.~Ebrero~(SRON),
	M.~Eckart~(NASA/GSFC),
	K.~Hagino~(JAXA),
	K.~S.~Long~(STScI),
	J.~Miller~(University~of~Michigan),
	T.~Miyazawa~(Nagoya~University),
	S.~Paltani~(Universit\`{e} de Gen\'{e}ve),
	C.~Reynolds~(University~of~Maryland),
	C.~Ricci~(Kyoto~University),
	H.~Sameshima~(JAXA),
	H.~Seta~(Saitama~University),
	Y.~Ueda~(Kyoto~University), and
	M.~Urry~(Yale~Unversity)
}
\MakeWhitePaperTitle

\begin{abstract}
In this white paper we describe the prospects for {\it ASTRO-H} for the study
of outflows from active galactic nuclei. The most important breakthroughs in
this field are expected to arise from the high spectral resolution and
sensitivity in the Fe-K band, combined with broad-band sensitivity over the full
X-ray band and spectral capabilities also at lower energies. The sensitivity in
the Fe-K region allows to extend the absorption measure distribution of the
outflow out to the highest ionisation states accessible, where observations with
current X-ray missions indicate that most of the outflowing gas is to be found.
Due to the high-resolution and sensitivity it will also be able to give the
definitive proof for the existence of ultra-fast outflows, and if so,
characterise their physical properties in great detail. These ultra-fast
outflows carry very large amounts of energy and momentum, and are of fundamental
importance for feedback studies. We show how the {\it ASTRO-H} observations
in general can help to constrain numerical models for outflows. The link to
reflection and emission processes is also discussed, as well as the possible
relation between outflows and relativistic emission lines. Finally, we discuss
the prospects for other related categories of objects like BAL quasars,
partially covered sources and Compton thick outflows.
\end{abstract}

\maketitle
\clearpage

\tableofcontents
\clearpage

\section{High-ionisation gas in ``normal'' outflows}

\subsection{Background and Previous Studies}

It has been recognised over the past years that AGN play a fundamental role in a
broad range of astrophysical processes. There is a tight relation between the
growth of the black holes and the evolution of the host galaxy, and feedback
processes between outflows from the black hole surroundings with their
environment may regulate this. The strongest feedback occurs for the
supermassive black holes at the centres of giant cD galaxies in cluster centres,
where the effects of the AGN activity in its ``radio mode'' can be felt up to
Mpc-scale distances. These cases are treated in detail in the cluster sections
of this white paper.

Here we focus on the outflows from ``normal'' AGN in their quasar mode. The
presence of outflows was first discovered from optical and UV spectra of AGN.
Evidence for photo-ionised gas surrounding the nucleus and giving rise to X-ray
absorption as visible through absorption edges of in particular O~VII and O~VIII
was first inferred from {\it ROSAT} PSPC data, although the credit of the
first discovery goes to \citet{halpern84}, who inferred the presence of
photo-ionised gas in MR~2251--178 using {\it Einstein} MPC data. With the
availability of CCDs through the launch of {\it ASCA}, the absorption edges
from the so-called warm absorbers were readily discovered and analysed for large
samples of AGN \citep[see][for an example]{reynolds97}. 

Subsequently, when the first X-ray grating spectrometers were launched
({\it Chandra} with the High Energy Transmission Grating Spectrometer, HETG,
and Low Energy Transmission Grating Spectrometer, LETG, and {\it XMM-Newton}
with the Reflection Grating Spectrometers, RGS), it was found that also line
absorption plays an important role, and that the warm absorbers show blueshifts,
indicating that they are outflowing. This was already clear from the first
high-resolution X-ray spectrum, taken with the LETG of the Seyfert galaxy
NGC~5548 \citep{kaastra00}, and confirmed soon after in NGC~3783
\citep{kaspi00}.

Based on a large number of papers that appeared in the literature since then,
the following picture arises. Outflows are seen in X-rays in about half of all
the type 1 sources, which is usually being interpreted that they cover about
half of the available solid angle of the ionisation cone. This is not
necessarily true if the outflow has significant bending due to e.g. ongoing
acceleration. Observations and modeling of UV lines in emission shows that the
column density rises for increasing inclination angle of the AGN
\citep{fischer12}, and showing a smooth transition towards type 2 sources,
indicating that the outflows may be stronger near the torus than the AGN axis.

Outflow velocities usually range between almost zero, to common values of a few
hundred km\,s$^{-1}$ up to a few 1000~km\,s$^{-1}$ in some cases (ultra-fast
outflows are discussed later). The number of components seen ranges between one
up to several, and usually the UV spectra have the best resolving power to
separate the different dynamical components, but they are by nature limited to
the lowly-ionised gas. In many cases, however, there is a clear connection
between the UV absorption components and the X-ray absorption components.

When we consider the ionisation structure of the outflow, already the first
LETG spectrum of an AGN outflow showed evidence for at least three ionisation
components \citep{kaastra02}. With more data available, the debate was arising
whether the absorption measure distribution (AMD), the function describing the
column density $N_{\rm H}$ as a function of ionisation parameter $\xi$, is
continuous or not. One point of view is that the AMD is power-law like over
three orders of magnitude in $\xi$ \citep{steenbrugge05}. Another point of view
is that the AMD has discrete components and that these components are in
pressure equilibrium \citep{krongold03}. Recently, very deep observations of
Mrk~509 showed that the AMD indeed has discrete components, but these are not in
pressure equilibrium \citep{detmers11}. See also Figure~\ref{fig:509stab}.

Different origins for the outflow have been proposed, like accretion disk winds,
thermal evaporation from the torus, or extended ionisation cones. These models
all imply very different effects for the impact of the outflow on the
environment, the feedback process. First, outflows launched close to the black
hole (like accretion disk winds) have to overcome a larger gravitational
potential to escape, hence should have much larger outflow velocities $v$, from
thousands of km/s to significant fractions of the speed of light. Furthermore,
outflows at larger distances $r$ from the black hole (like e.g. torus winds)
have a larger impact on the surroundings, because we have for the mass loss rate
$\dot M$: 
\begin{equation} 
\dot M / \Omega = N_{\rm H} m_p r \textrm{v}, \label{eqn:mdot}
\end{equation} 
with $\Omega$ the solid angle subtended by the outflow, $m_p$ the proton mass
and $N_{\rm H}$ the total column density (dominated by hydrogen). The kinetic
luminosity $L_K$ is simply given by $\frac{1}{2}\dot M \textrm{v}^2$. Thus, the
fastest winds generally dump most energy into their surroundings, and therefore
they are of most interest for studies of cosmic feedback.

Empirically, it is not easy to determine the location of the outflow, because
the spectrum only yields the physical state (from the ions that are present and
line broadening), the total column density (from the depth of the lines) and the
outflow velocity (from the Doppler shift of the lines). However, because the
ionisation parameter $\xi$ is given by
\begin{equation}
\xi = L/nr^2
\end{equation}
with $L$ the ionising luminosity, $n$ the hydrogen density and $r$ the distance
of the gas from the central black hole, a density measurement would yield
directly the distance, because both $L$ and $\xi$ can be deduced directly from
the spectrum. $\xi$ is usually expressed in units of erg\,s$^{-1}$\,cm or
$10^{-9}$~Wm, which is used throughout this paper. The density can sometimes be
obtained using UV absorption lines from meta-stable levels, for the lowly
ionised gas components, or through measuring the response of the outflow to
variations in the ionising continuum (i.e., measuring the recombination time
scale which is proportional to $1/n$). The latter method requires monitoring of
a source at regular intervals, and has been applied successfully to Mrk~509
\citep{kaastra12}. In the latter case it was found that the dominant X-ray
absorption components are located at a distance of a few to a few tens of pc
from the nucleus, near the inner parts of the torus.

Interestingly, several AGN show a tendency of increasing outflow velocity as a
function of ionisation parameter \citep{steenbrugge05,tombesi13}. A major
uncertainty, however, is how this trend continues to higher ionisation parameter
(say $\log\xi >3$), and in addition, how the AMD behaves at high ionisation
parameter. At lower ionisation parameter, the AMD, defined as 
\begin{equation} 
\textrm{AMD} = \xi {\rm d}N_{\rm H}/{\rm d}\xi 
\end{equation} 
increases approximately as a power law function of $\log\xi$ (at the discrete
$\xi$-values where it is non-zero), with power-law index typically 0.0 -- 0.4,
\citep[cf.][]{behar09}. Thus, there is more gas at higher ionisation parameter,
unless of course there would be a natural cut-off to the AMD.

\subsection{Prospects \& Strategy}

The increase of the AMD towards higher ionisation parameter, as well as the
tendency of higher outflow velocities at higher ionisation parameter, offer
unique prospects for {\it ASTRO-H} to study AGN outflows, because this may
imply that the dynamically most important part of the outflow occurs at these
high values of $\xi$. Observationally, the highest ionised gas components are
best studied by measuring the K-shell transitions of Fe~XXV and Fe~XXVI.

We note here that the Fe-K band also contains strong absorption features from
transitions of Li-like Fe and higher iso-electronic sequences, and
{\it ASTRO-H} measurements will definitely make use of these last types of
transitions. However, the same ions have also transitions in the L-band near
1~keV, also detectable by SXS, but in that energy band the grating spectrometers
of {\it Chandra} and {\it XMM-Newton} are getting more sensitive and have
already explored quite well parameter space (see the introduction before).
Therefore, here we focus uniquely on the prospects for discoveries based on
Fe~XXV and Fe~XXVI. However, we note here that in the same spectral region, for
very well exposed spectra, also the equivalent nickel lines will become
accessible.

\begin{figure}
\begin{center}
\includegraphics[width=0.65\hsize]{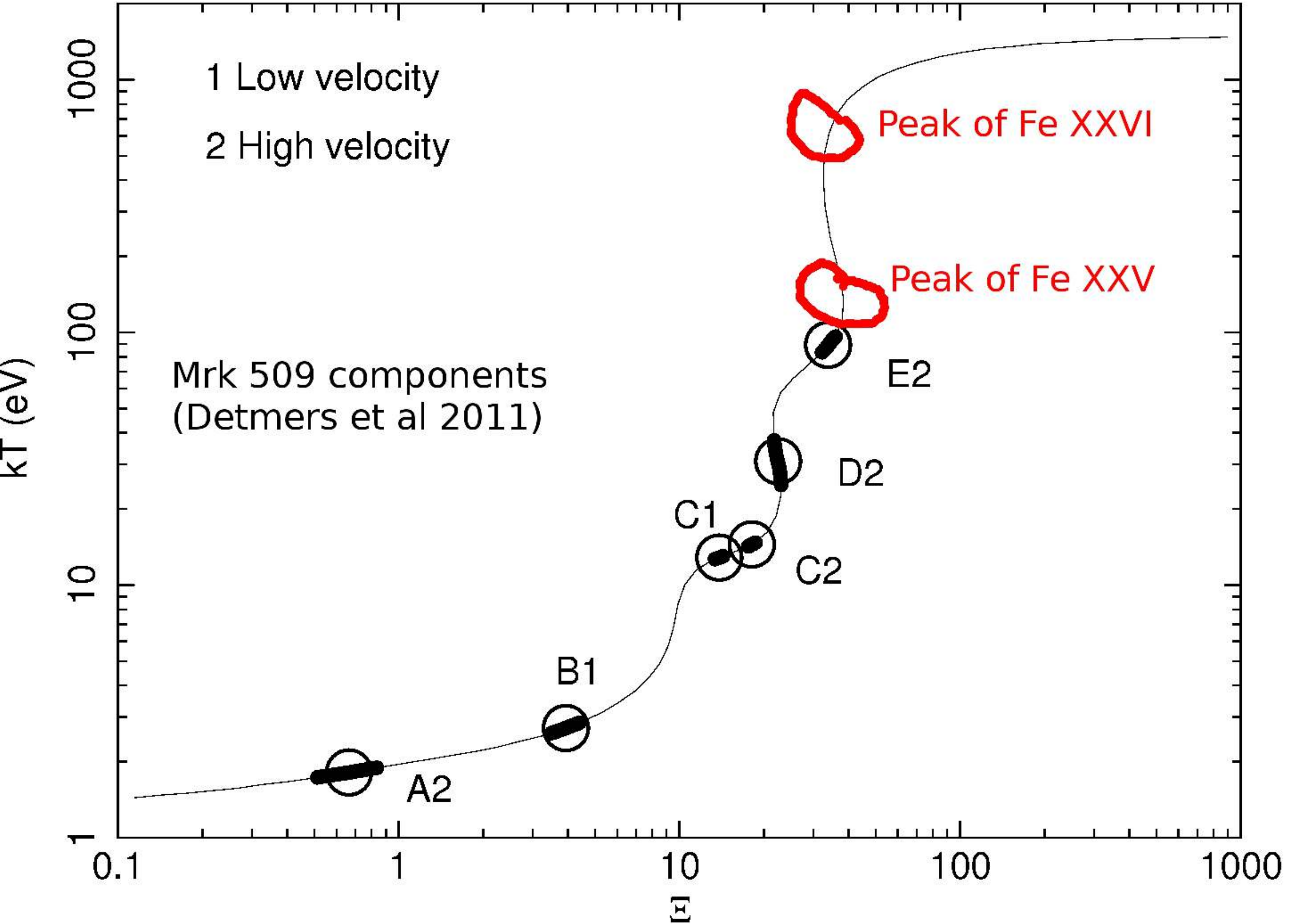}
\end{center}
\caption{Photoionisation equilibrium curve for Mrk~509, based on
\protect\citet{detmers11}. Components are in pressure equilibrium when they
share the same value of $\Xi\equiv\xi/4\pi ckT$. Stable equilibrium only exists
on parts of the curve with positive slope. The discrete components found by
Detmers et al. are indicated by circles and labels corresponding to the
designation in that paper, and the thick parts on the curve correspond to the 
measured uncertainty in the ionisation parameter $\xi$ for these components. In
red we indicate at which parts of the curve Fe~XXV and Fe~XXVI are expected to
peak.}
\label{fig:509stab}
\end{figure}

We first consider the stability curve of photo-ionisation equilibrium and the
different discrete phases. Figure~\ref{fig:509stab} shows the photo-ionisation
equilibrium curve for Mrk~509 with the discrete components found by
\citet{detmers11}. It is seen that we expect Fe~XXV and Fe~XXVI to peak around
the position of the last unstable branch, and in the latter case already on the
stable Compton branch. Measuring the column density, outflow velocity,
turbulence and other parameters for these ions therefore will provide us unique
insight into the dynamically most important component of the outflow. Even a
non-detection would give extremely useful upper limits. 

Up to now, there have been two clear cases of absorption line detections in the
Fe-K band. One case is NGC~3783, where \cite{reeves04} found evidence
($4.5\sigma$) for the Fe~XXV 1s--2p absorption line in a deep
{\it XMM-Newton} EPIC spectrum. There was evidence for variability on a
$10^5$~s time scale. NGC~3783 is an AGN with one of the highest outflow column
densities, explaining why a limited detection with a CCD camera was possible.
The other case is the deep 520~ks {\it Chandra} HETG spectrum of
MCG\,$-$6-30-15, showing H-like and He-like Fe absorption lines outflowing at
2000~km\,s$^{-1}$ at a very high column density of $10^{23.2}$~cm$^{-2}$ at
$\log\xi=3.6$ \citep{young05}.

As we will show, it is not very hard to get line detections with SXS, but in
order to constrain the physical parameters we need better quality spectra hence
longer exposure times. This will be elaborated in the next section. It is
important to get a well-justified sample of sources spanning a broad range of
parameters, from low mass narrow-line Seyfert galaxies to high-mass quasars, and
preferably also with other different parameters such as column density of the
outflow, different ionising spectral energy distributions, etc.

Before proceeding, however, we first make some basic estimates that help to
prepare the source selection.

\subsubsection{Absorption line detection}

Here we provide a few simple estimates of what column densities can be detected
with SXS in the Fe-K band. Starting point is a power-law continuum with
effective photon index 1.7 and 2--10 keV flux $F$ of 5 units (one unit is here
$10^{-14}$~W\,m$^{-2}$ or $10^{-11}$~erg\,cm$^{-2}$\,s$^{-1}$). This is the
typical flux of one of the brighter Seyfert galaxies. We consider a canonical
exposure time $t$ of 100~ks, and the SXS effective area $A(E)$, with spectral
resolution $\Delta E = 5$~eV. Let the equivalent width of the line be $W$. If
the line is weak, $W\ll \Delta E$, and if the line is strong, $W\gg \Delta E$. 

For any line, the number of continuum counts $N_{\rm c}$ acting as
``background''
for line detection is:
\begin{equation}
N_{\rm c} = A(E)\, t\, F(E)\, W_{\rm{eff}},
\end{equation}
where $F(E)$ is photon flux in photons\,area$^{-1}$\,time$^{-1}$\,energy$^{-1}$,
and
\begin{equation}
\textrm{Weak line:}\ \ W_{\rm{eff}} = \Delta E,
\end{equation}
\begin{equation}
\textrm{Strong line:}\ \ W_{\rm{eff}} = W.
\end{equation}

The number of line photons $N_{\rm l}$ taken away from the continuum is 
\begin{equation} 
N_{\rm l} = A(E)\, t\, F(E)\, W.
\end{equation}

The signal to noise ratio S/N of the detection (``number of sigma's'' or $n\sigma$
of the detection) is given by
\begin{equation}
\textrm{weak line:}\ \ \textrm{S/N} = 
N_{\rm l} / \sqrt{N_{\rm c}} = \sqrt{A(E)\, t\, F(E)}\,\, W / \sqrt{\Delta E}.
\end{equation}
For a strong absorption line, it is essentially provided by the uncertainty in
the number of continuum photons that are absorbed (assuming a ``black'' line):
\begin{equation}
\textrm{strong line:}\ \ \textrm{S/N} = 
\sqrt{N_{\rm c}} = \sqrt{A(E)\, t\, F(E)}\,\, \sqrt{W}.
\label{eqn:strongline}
\end{equation}
These results can be generalised to
\begin{equation}
\textrm{S/N} = \sqrt{A(E)\, t\, F(E)}\,\, \frac{W}{\sqrt{\max (\Delta E, W )}}.
\end{equation}

\subsubsection{Practical example}

\begin{figure}[!tip]
\begin{center}
\resizebox{0.33\hsize}{!}{\includegraphics[angle=-90]{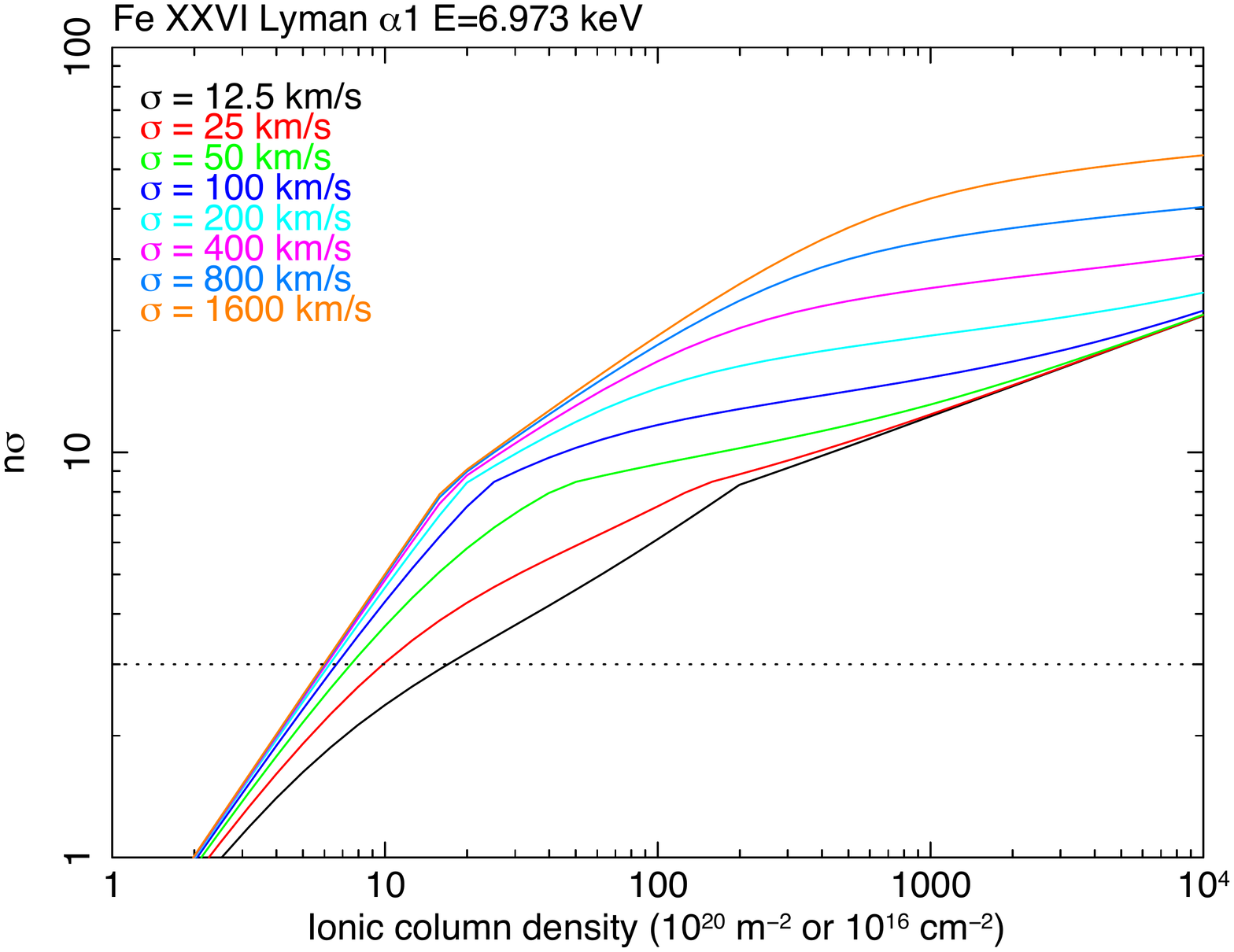}}
\resizebox{0.33\hsize}{!}{\includegraphics[angle=-90]{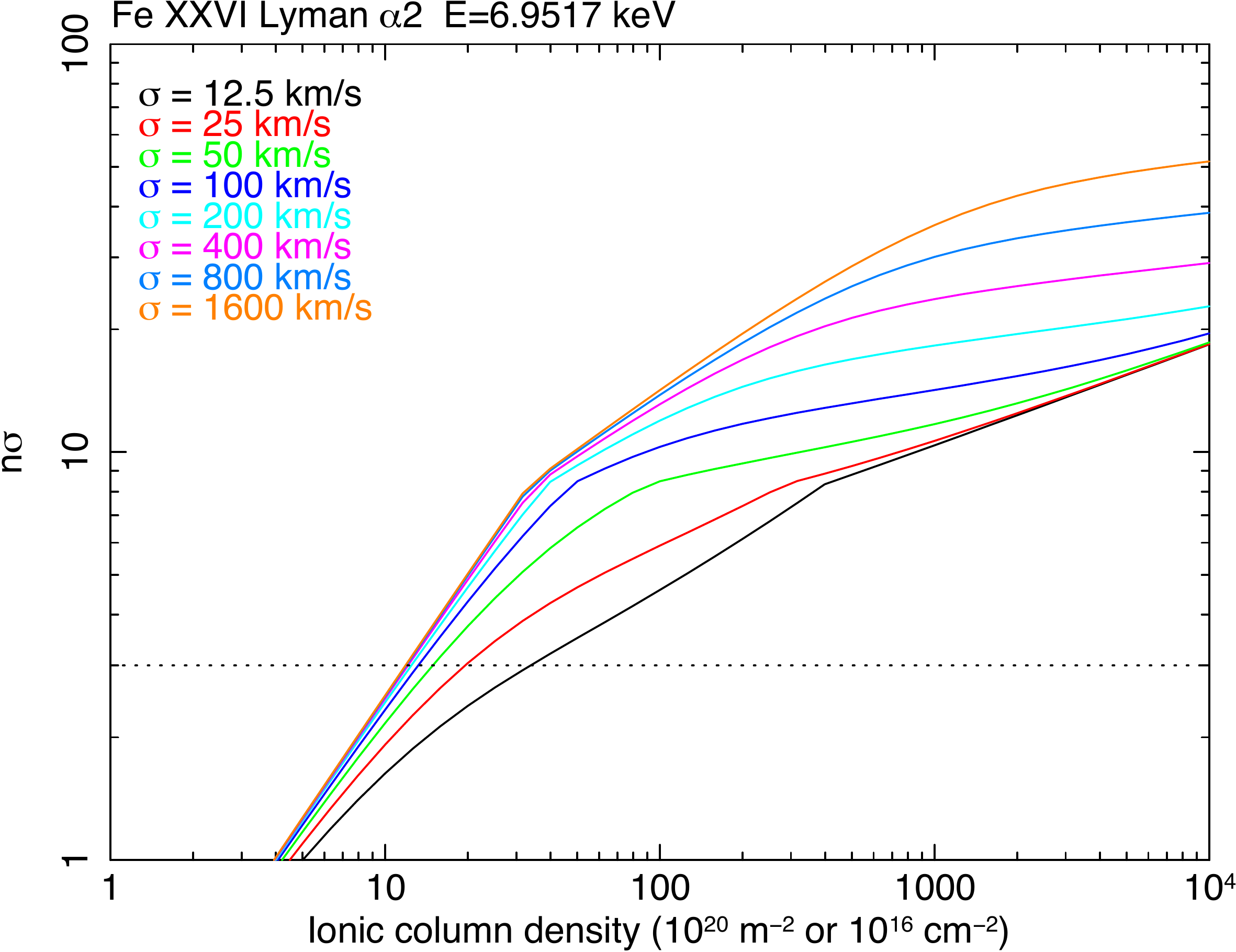}}\\
\resizebox{0.33\hsize}{!}{\includegraphics[angle=-90]{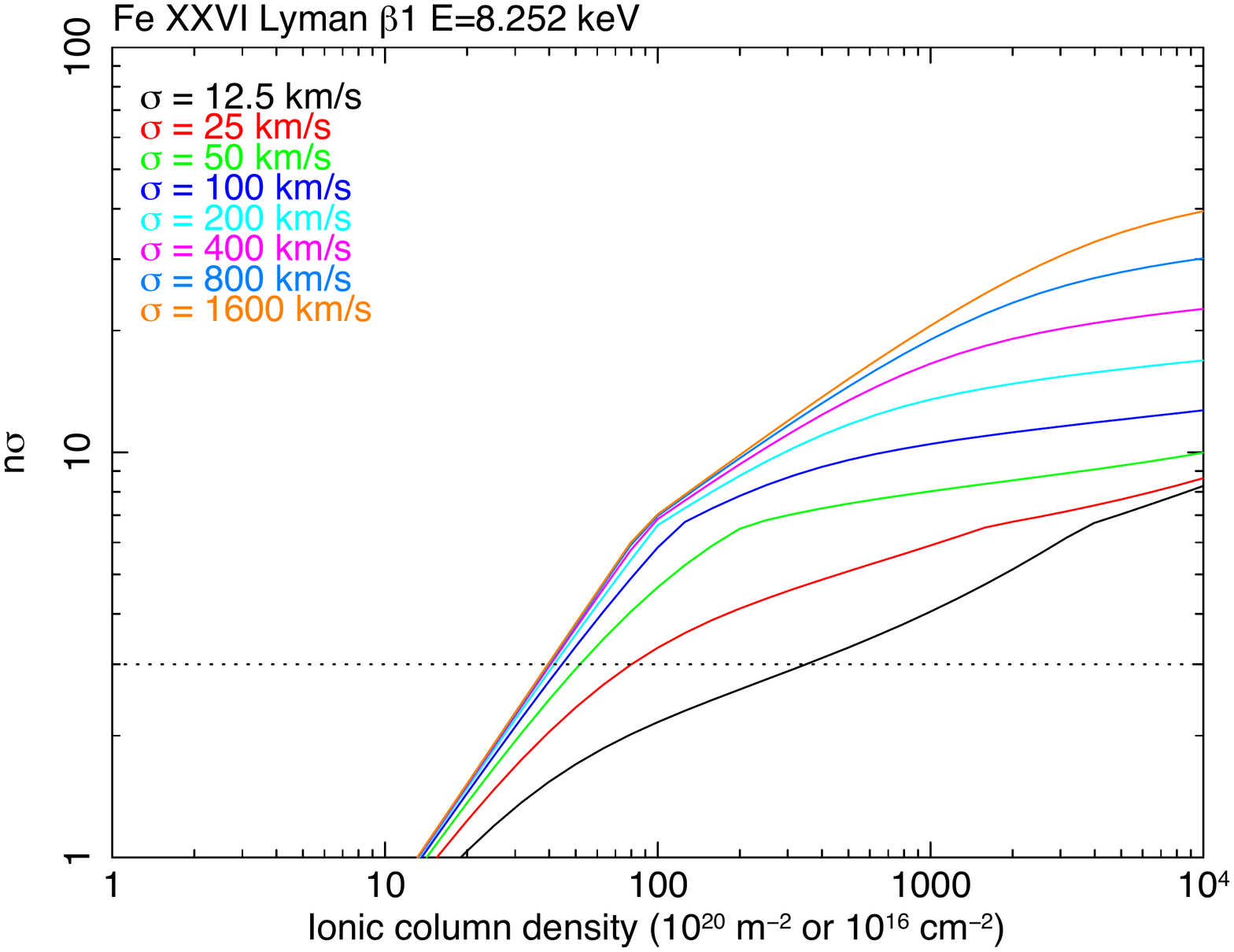}}
\resizebox{0.33\hsize}{!}{\includegraphics[angle=-90]{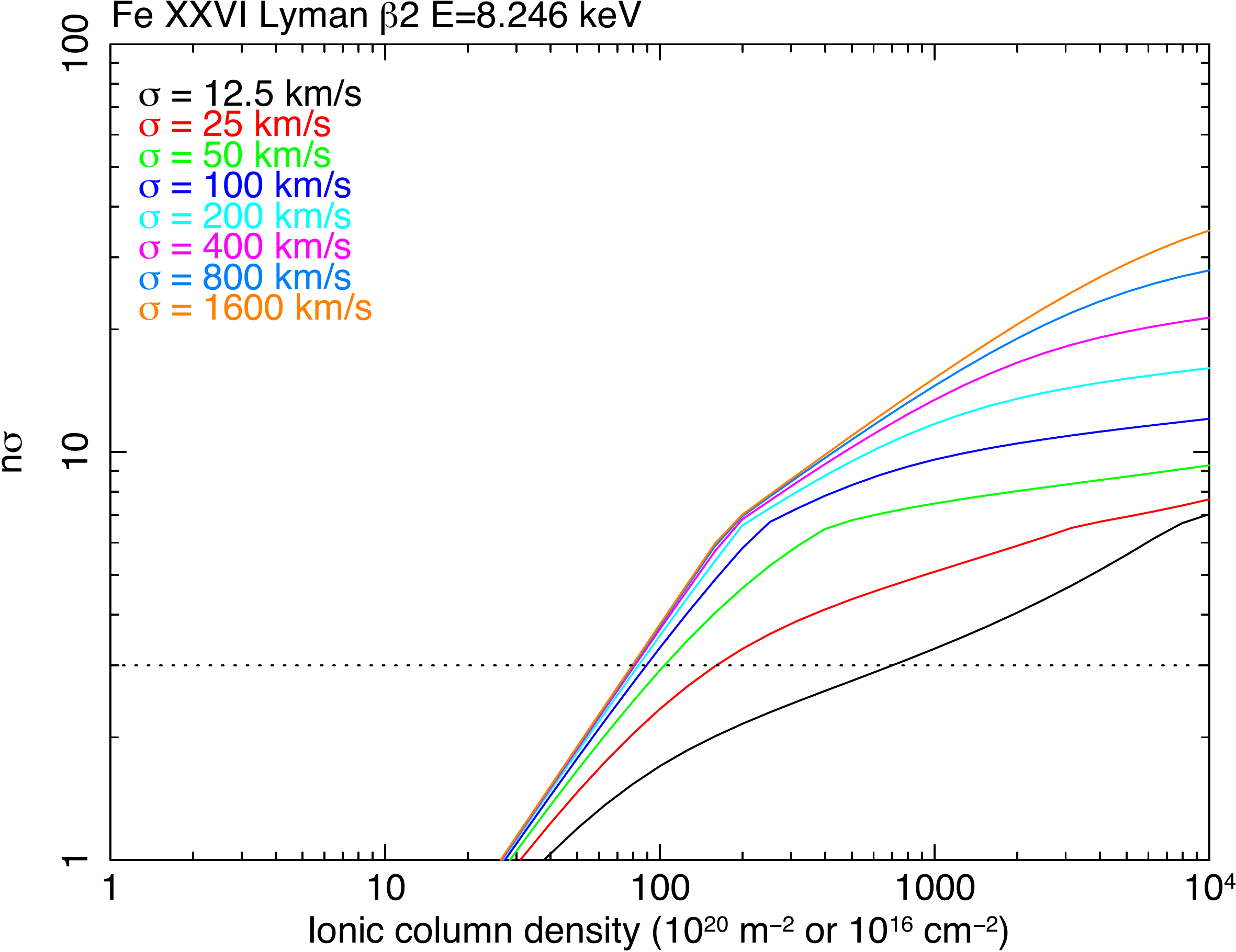}}\\
\resizebox{0.33\hsize}{!}{\includegraphics[angle=-90]{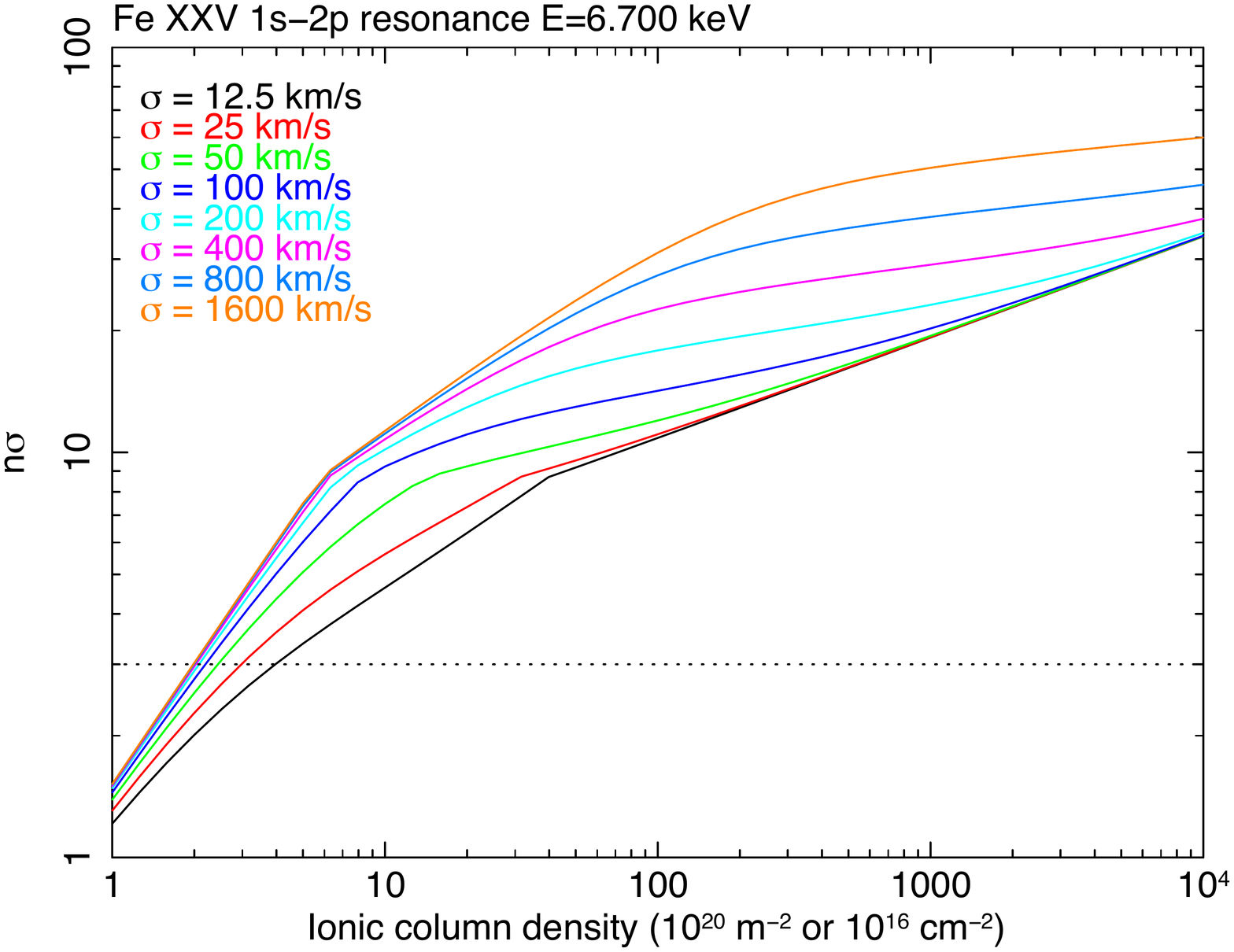}}
\resizebox{0.33\hsize}{!}{\includegraphics[angle=-90]{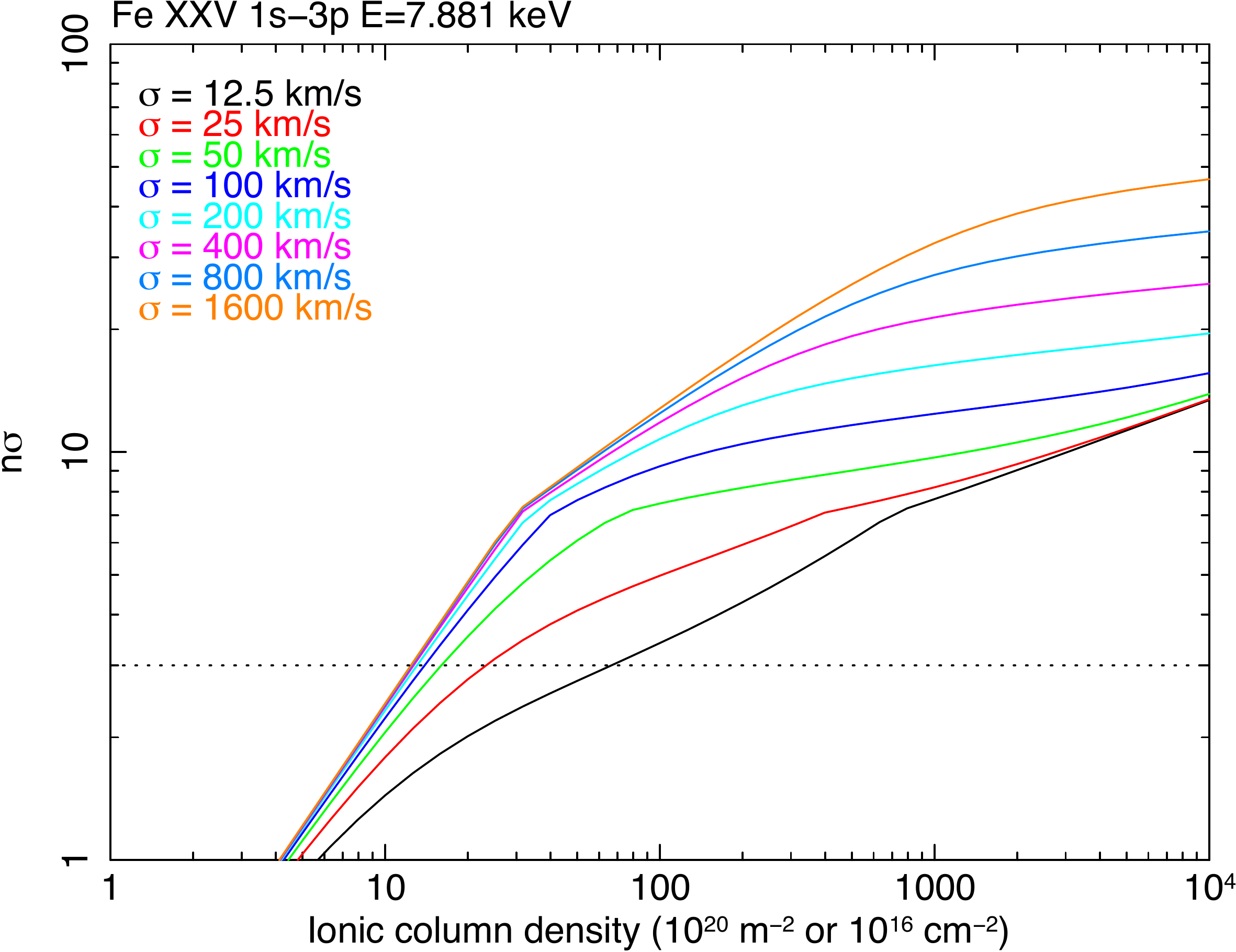}}\\
\resizebox{0.33\hsize}{!}{\includegraphics[angle=-90]{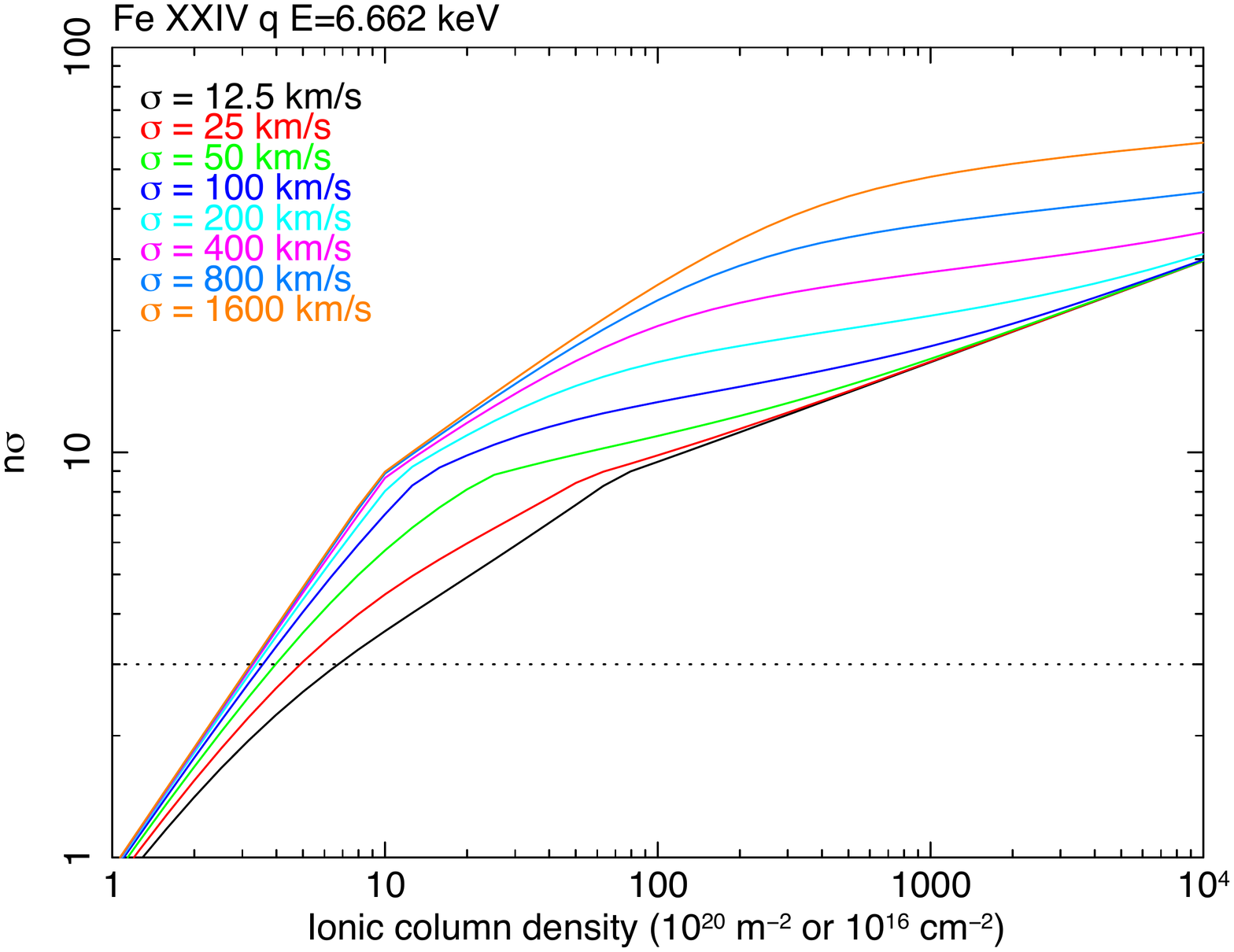}}
\resizebox{0.33\hsize}{!}{\includegraphics[angle=-90]{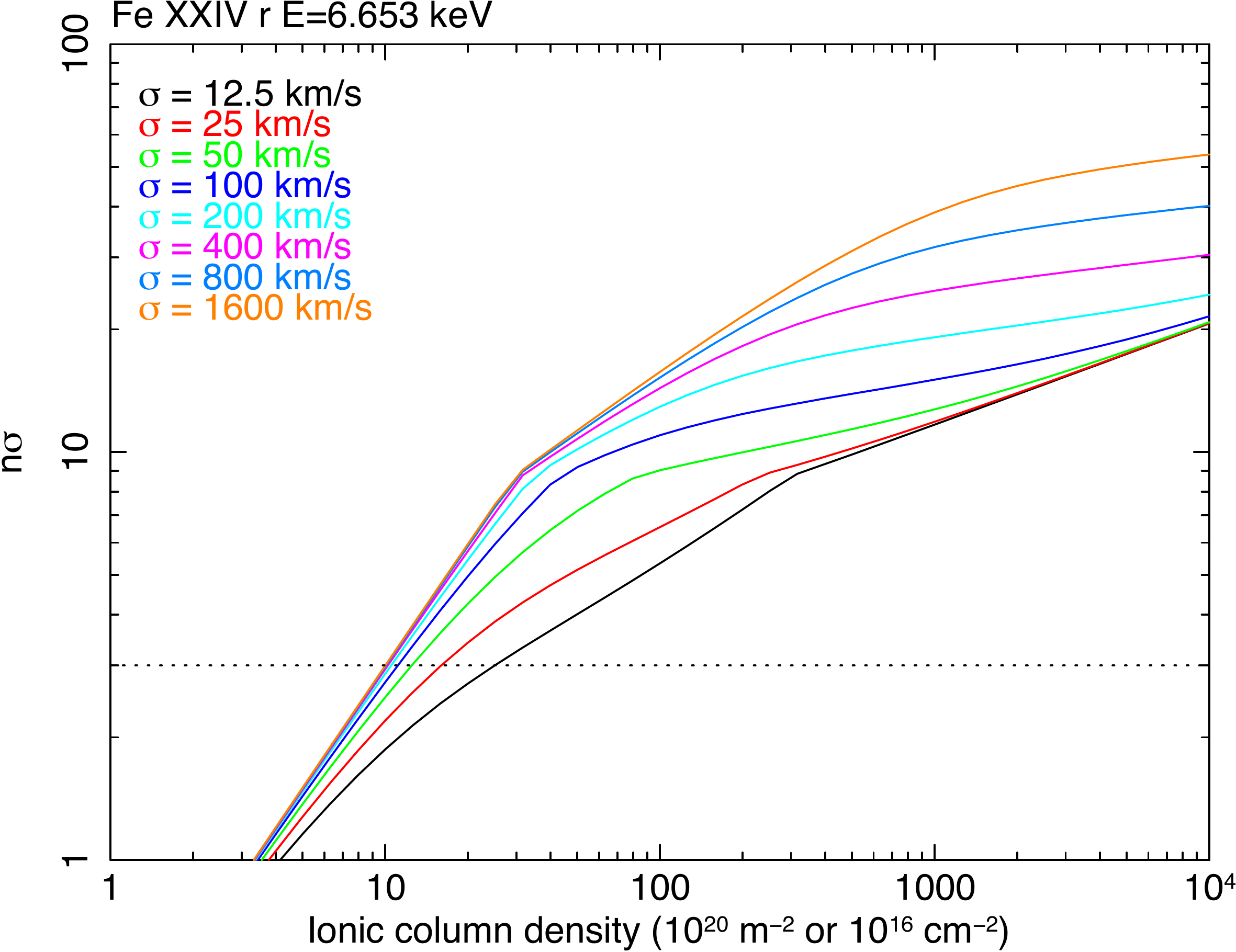}}\\
\resizebox{0.33\hsize}{!}{\includegraphics[angle=-90]{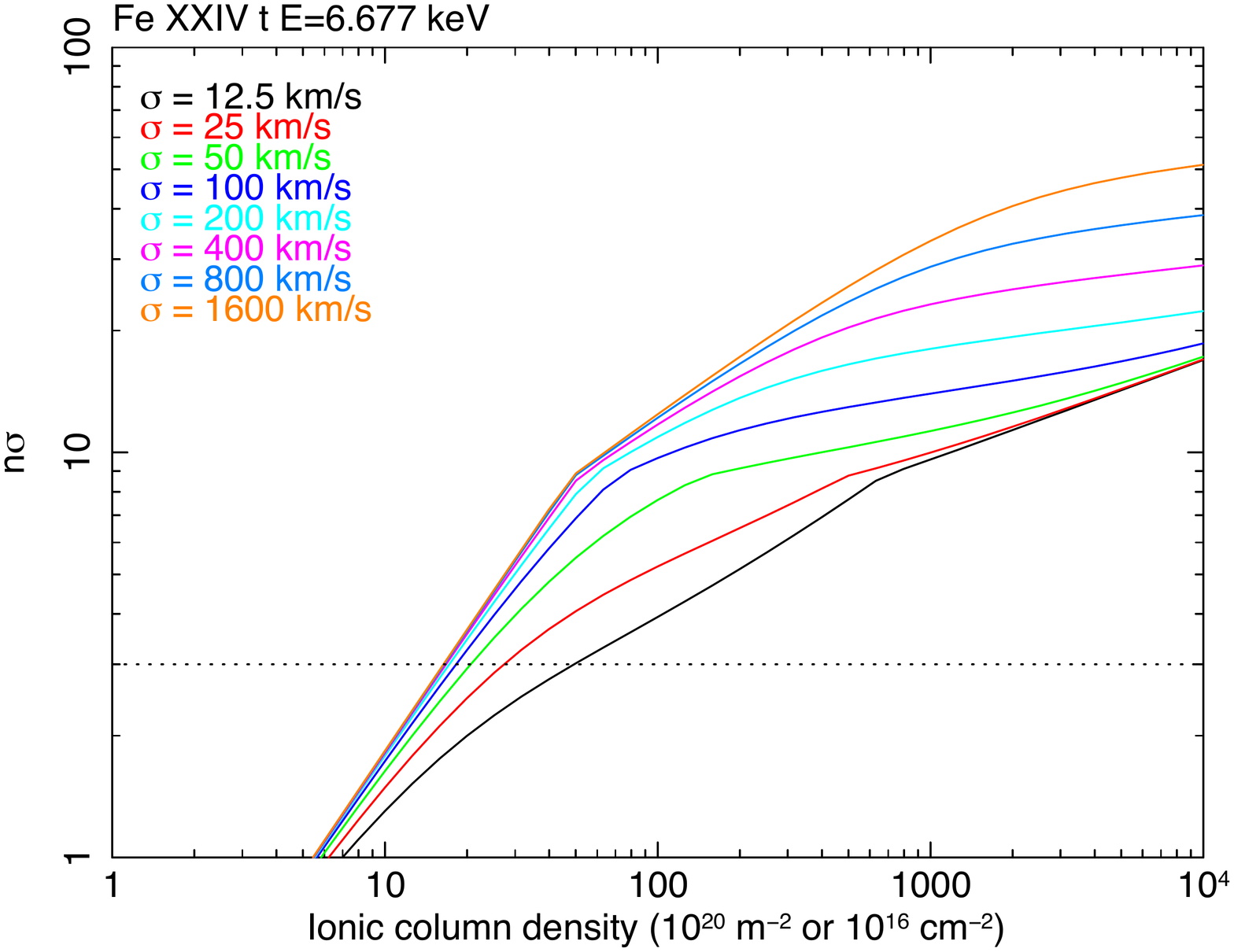}}
\resizebox{0.33\hsize}{!}{\includegraphics[angle=-90]{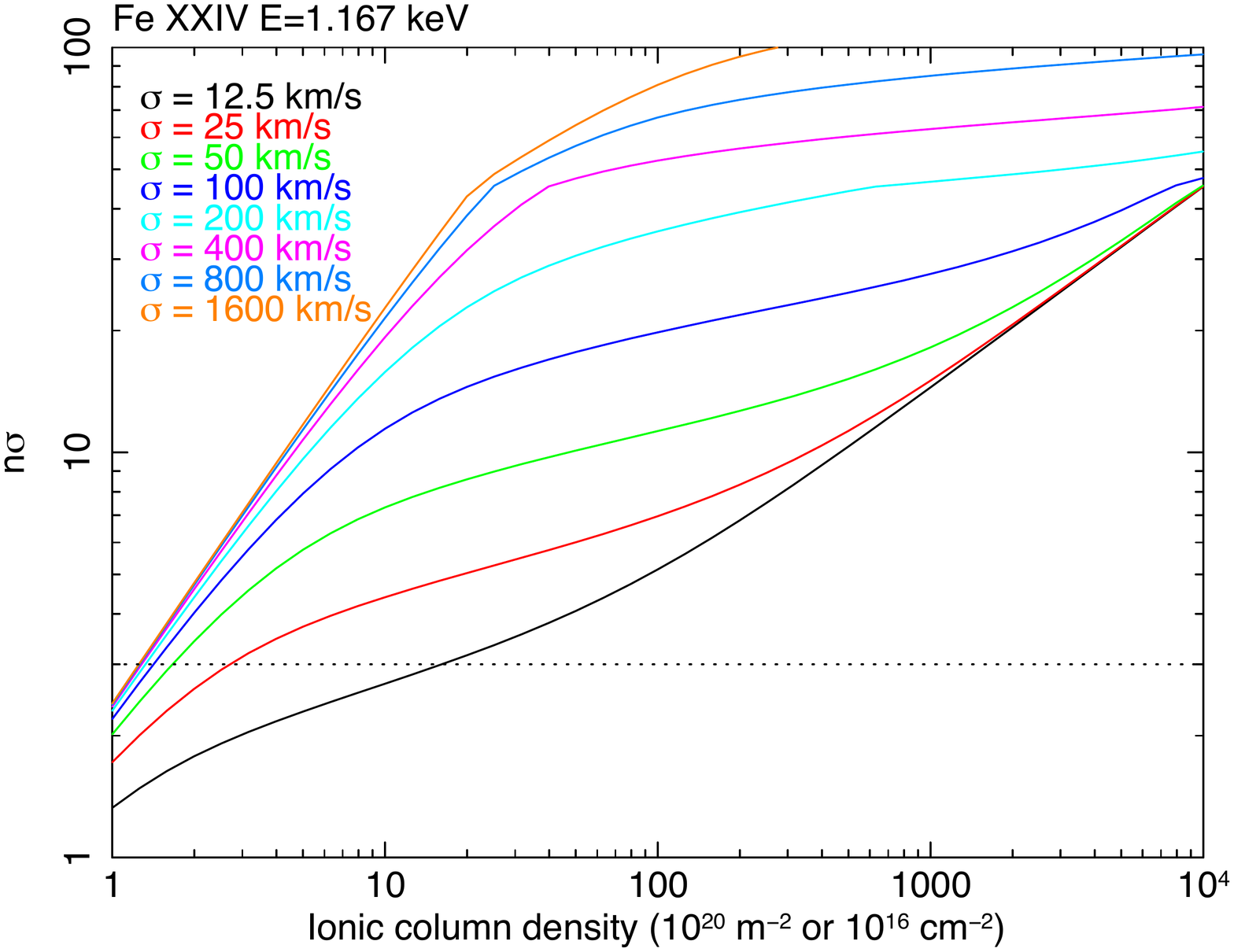}}
\end{center}
\caption{Detection significances ($n\sigma$ is S/N) for a 100~ks exposure on a
typical bright AGN with parameters as described in the text, for different
Gaussian velocity broadening as indicated. The dotted horizontal lines indicate
the 3$\sigma$ detection limits.} 
\label{fig:ews}
\end{figure}

The above formalism has been applied to absorption lines from a few ions: the Fe
XXVI Lyman $\alpha$ and $\beta$ lines (1s--2p and 1s--3p), the Fe XXV 1s--2p and
1s--3p lines, the Fe XXIV q, r and t lines, and the strongest Fe XXIV line in
the Fe-L band. Figure~\ref{fig:ews} shows the significance of line detection for
different values of the Gaussian velocity dispersion $\sigma$ as indicated.

For Solar abundances, the maximum concentrations that the ions Fe XXIV, Fe XXV
and Fe XXVI relative to hydrogen can reach, are $6\times 10^{-6}$, $1.7\times
10^{-5}$, and $1.5\times 10^{-5}$ respectively.

Thus, for instance, to detect the strong Fe XXV resonance line at 6.7 keV: from
the figure we see that the $3\sigma$ detection limit is reached for an ionic
column density of about $2-4 \times 10^{20}$~m$^{-2}$, depending upon the
velocity dispersion, and thus the minimum equivalent hydrogen column density
(for Solar abundances) is $1-2 \times 10^{25}$~m$^{-2}$ ($1-2 \times
10^{21}$~cm$^{-2}$). Such columns are not very large and easily detectable.
However, note the scaling with source flux: for a ten times weaker source, the
minimum detectable column density may be two orders of magnitude larger.

\subsection{Targets \& Feasibility}

We have selected targets starting from a few review papers. We have chosen here
the reviews by \citet{blustin05}, \citet{mckernan07} and \citet{tombesi12}. From
these, we took 23 sources with measured $\xi$ and column density. When multiple
components were given, we took the one with the highest ionisation parameter. We
estimated equivalent hydrogen columns for Fe~XXV and Fe~XXVI by assuming they
peak at $\log\xi$ values of 3.7 and 4.1, respectively (these values are
appropriate for Mrk~509). We then scaled column densities to these values by
scaling with $\xi^{0.2}$, a typical value found by \citet{behar09}. Median 2--10
keV fluxes were taken from two sources: the \citet{tombesi10a} paper and a
private list of 2--10~keV fluxes from various satellites in the
pre-{\it ASCA} era.

For each of these sources, we simulated the spectrum using a power law with
photon index 1.7 and the appropriate 2--10~keV flux, with the absorption by
Fe~XXV and Fe~XXVI calculated through the \textsl{slab} model of SPEX, which
calculates the opacity for each individual ion. A uniform Gaussian velocity
broadening of $\sigma=$100~km\,s$^{-1}$ was adopted. We then simulated these
spectra for SXS using 500~ks exposure time and determined the accuracy of the
measured column densities. The relative uncertainties on the measured ionic
column densities were inverted to get the equivalent number of sigmas for the
significance. We found a dozen sources where a 5$\sigma$ detection for both ions
within 500~ks is possible.

One needs to take care, however, in selecting sources based on such simulations.
For instance, for MCG~8-11-11 it appears that the initial high column density
($N_{\rm H}=10^{22}$~cm$^{-2}$) found by \citet{matt06} from a fit to a 38~ks
{\it XMM-Newton}/EPIC spectrum, is not confirmed by a deeper {\it Suzaku}
observation and in fact may be 30 times lower \citep{bianchi10}. Also, a yet
unpublished 120~ks {\it Chandra} HETG spectrum shows no signs of strong
absorption lines.

In the next sections we make a few dedicated simulations for {\it ASTRO-H}
for some of the most promising sources.

\subsubsection{NGC~4151}

\begin{figure}[!htbp]
\begin{center}
\resizebox{0.43\hsize}{!}{\includegraphics[]{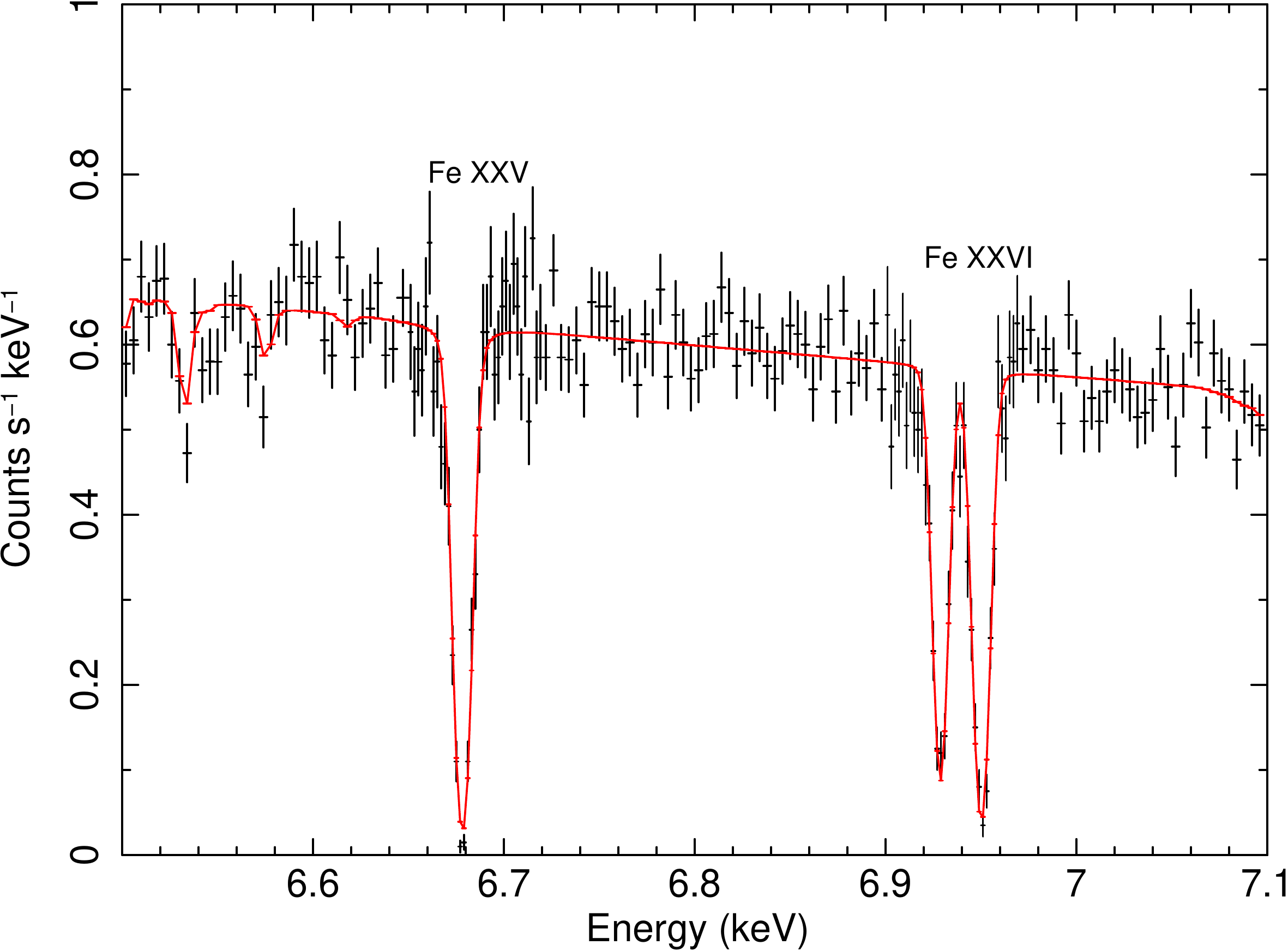}}
\resizebox{0.43\hsize}{!}{\includegraphics[]{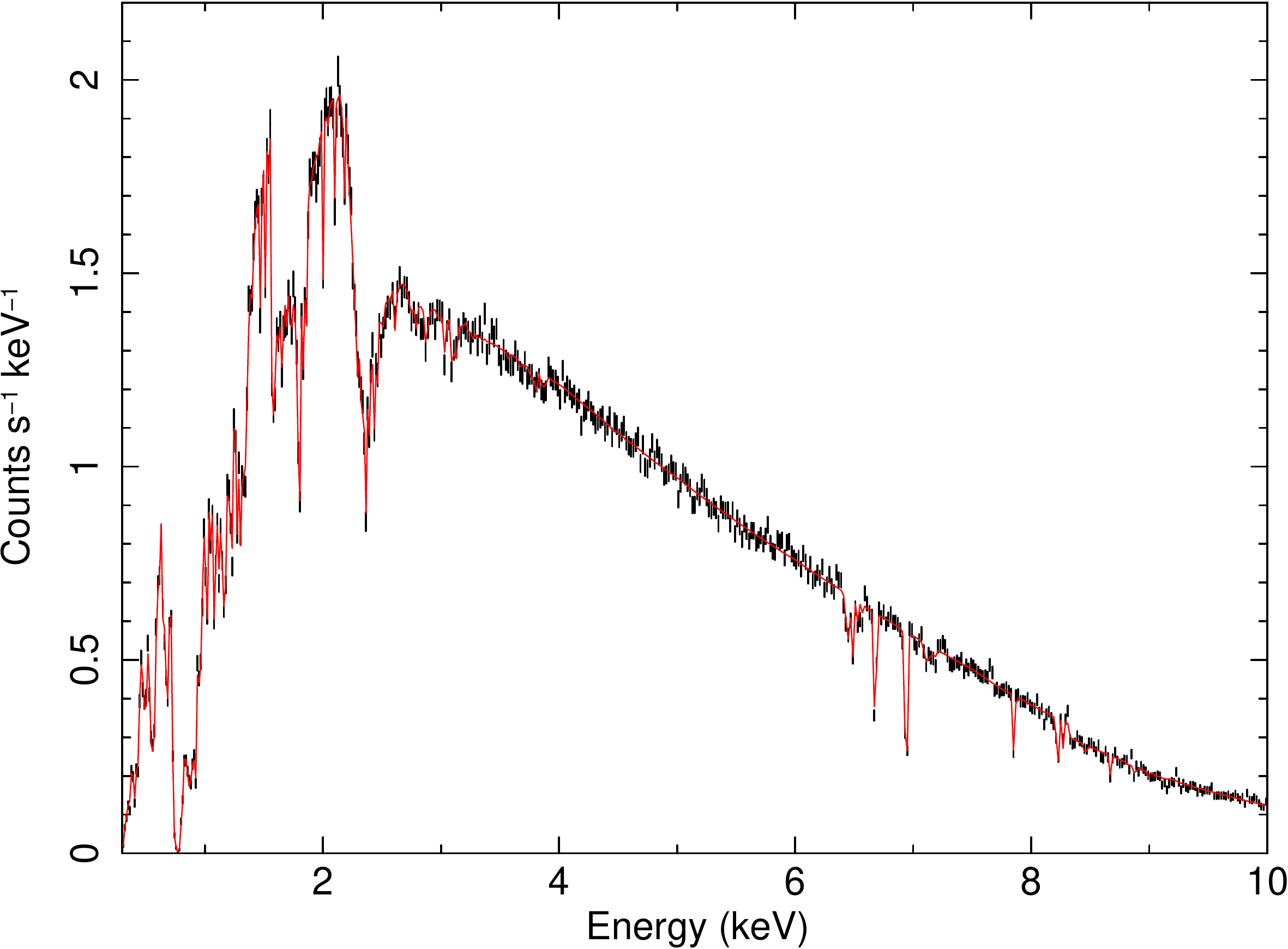}}
\end{center}
\caption{Simulated spectrum for NGC~4151 for 100~ks exposure time.}
\label{fig:ngc4151}
\end{figure}

We simulated the spectrum of NGC~4151 using the same flux condition of the 2002
HETG spectrum \citep{kraemer05}. We used their absorber model (adopting the
\textsl{xabs} model fo SPEX for the absorbers) and added two \textsl{slab}
components for the higher ionization gas. See Table~\ref{tab:4151sim} for the
parameters and Figure~\ref{fig:ngc4151} for the simulated spectrum.

\begin{table}[htdp]
\caption{Simulated parameters for NGC~4151. Column densities in cm$^{-2}$.}
\begin{center}
\begin{tabular}{lc}
\hline
\multicolumn{2}{c}{Simulated parameters}\\
\hline
Exposure & 100 ks \\
Flux (2--10 keV) & $1.9 \times 10^{-10}$ cgs \\
xabs 1 & $\log N_{H}=22.5$  ; $\log \xi=2.55$\\
xabs 2 & $\log N_{H}=22.46$  ; $\log \xi=1.23$\\
xabs 3 & $\log N_{H}=20.8$  ; $\log \xi=-0.17$\\
xabs 4 & $\log N_{H}=21.6$  ; $\log \xi=0.42$\\
xabs 5 & $ \log N_{H}=20.8$  ; $\log \xi=-0.09$\\
\hline
\multicolumn{2}{c}{Fitted parameters}\\
\hline
slab 1 & $\log N_{\rm{Fe\,XXV}}=17.94\pm0.02$  \\
slab 2 & $\log N_{\rm{Fe\,XXVI}}=17.88\pm0.04$  \\
\hline
\end{tabular}
\end{center}
\label{tab:4151sim}
\end{table}

Note that the flux during the 2002 {\it Chandra} HETG observation that we
used as a template is rather high. It is in most cases a factor of 2 dimmer, on
average. Hence, a twice as long exposure time (200~ks) would be needed for a
typical flux state.

\subsubsection{NGC~3783\label{sect:3783}}

\begin{figure}[!htbp]
\begin{center}
\resizebox{0.43\hsize}{!}{\includegraphics[]{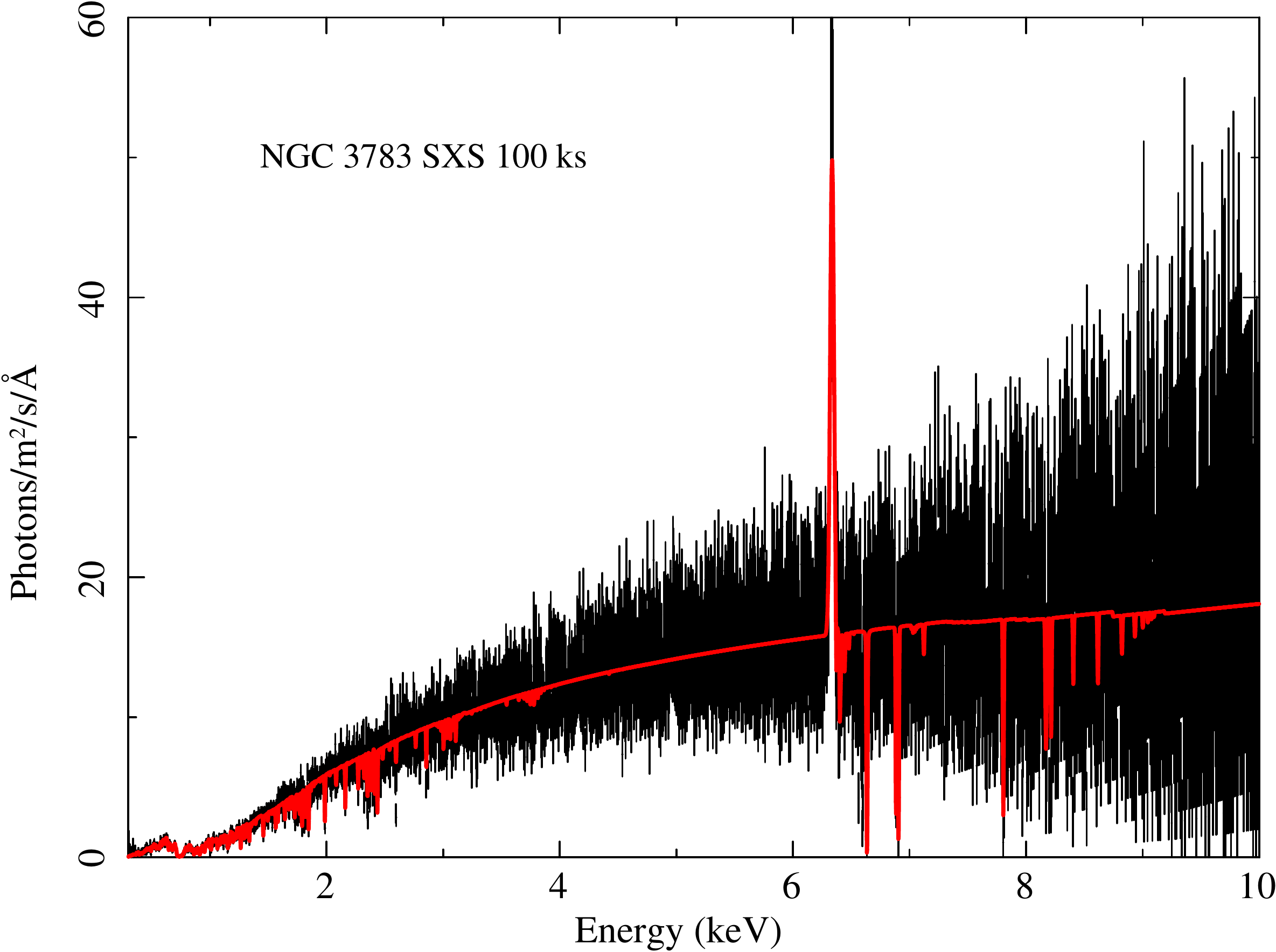}}
\resizebox{0.43\hsize}{!}{\includegraphics[]{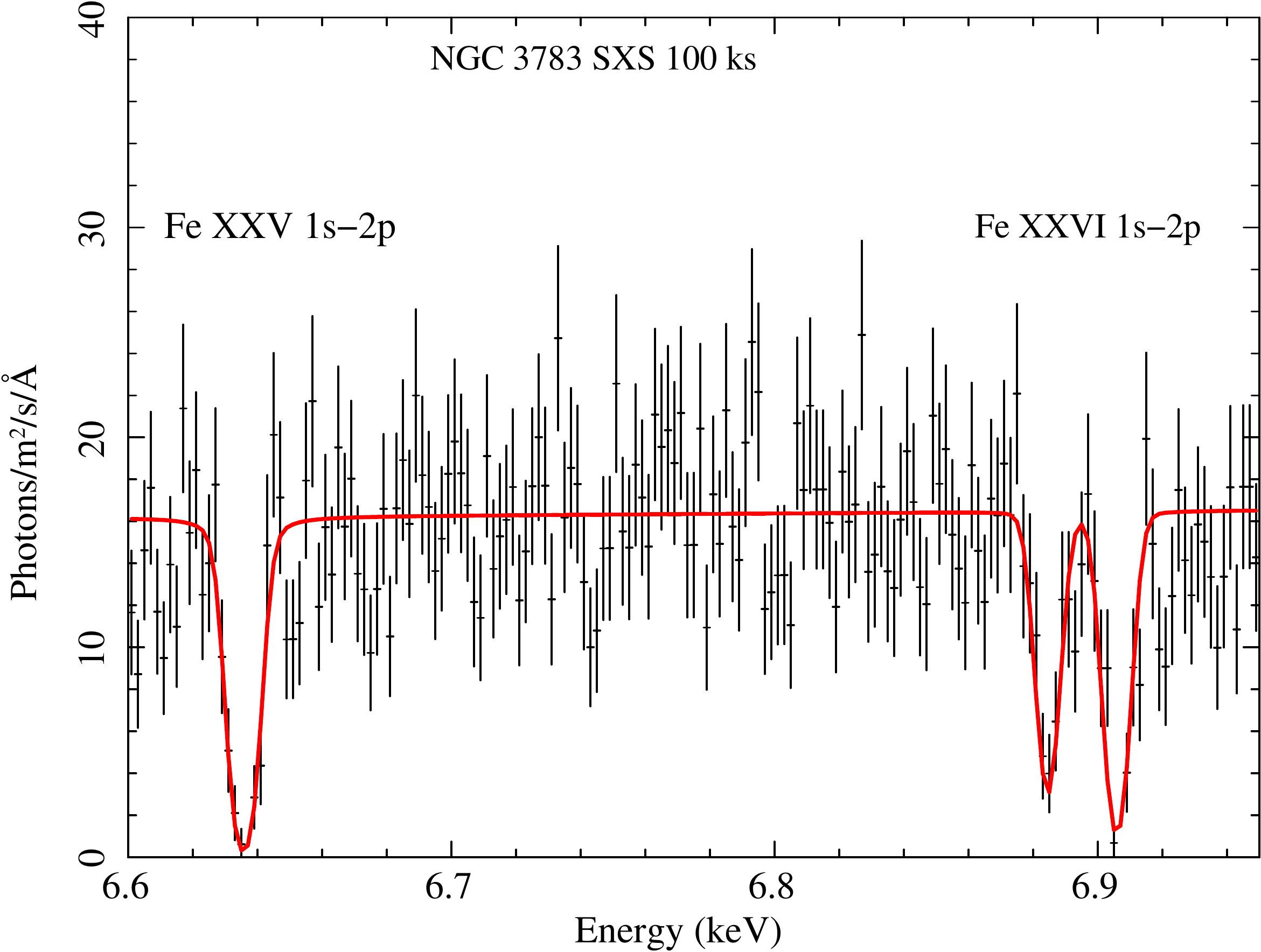}}
\end{center}
\caption{Simulated spectrum for NGC~3783 for 100~ks exposure time. The left
panel shows the full spectrum; due to the strong compression of this high
resolution spectrum along the horizontal axis, the error bars look larger than
they are: the right panel shows a blow-up near the iron lines and
demonstrates the true power of the SXS.}
\label{fig:ngc3783}
\end{figure}

The spectrum of NGC~3783 was simulated as follows (Figure~\ref{fig:ngc3783}). The
continuum parameters were adjusted in the way described earlier in this white
paper with a 2--10 keV flux of $5.2\times 10^{-11}$~cgs. The Fe K$\alpha$
emission line was modelled with a single Gaussian whose parameters match the
average result of the \citet{yaqoob05} analysis ($E_0 = 6.397$~keV, FWHM =
1700~km\,s$^{-1}$). The model includes 3 lower $\xi$ components, as described
in  \citet{netzer03}. For the simulation we added two more components with
corresponding ionization parameters of 3.7 and 4.1, respectively. Their column
densities are scaled from the highest value reported in \citet{netzer03} using
$\xi^{0.2}$. For an exposure time of 100~ks we can determine the Fe~XXV and
Fe~XXVI columns with an accuracy of 17\% and 19\%, respectively. The optical
depth of the strongest 1s--2p transitions at line center is 15.7 for Fe~XXV, and
5.0 for Fe~XXVI, and so the lines have a strong saturation. This means that even
if the column density is much lower, the lines still can be detected easily.

\subsubsection{NGC~5548\label{sect:5548}}

We re-analysed the 340~ks {\it Chandra} LETG data of the Seyfert 1 galaxy NGC
5548 using the radiation transfer code with the latest atomic data
(\texttt{CLOUDY} vc10; Ferland 2002) and the software package optimized for
high-resolution X-ray spectroscopy (\texttt{SPEX} v2.04; \citealt{kaastra96}).
Following the previous work \citep{steenbrugge05}, we fitted the spectrum with a
model consisting of a power-law plus a black body continuum with ionized
absorbers. We found that at least four distinct absorbers with different
ionization degrees are required to explain the absorption features. 
\citet{steenbrugge05} showed that the spectrum represents five absorption
components. We derived the ionization parameter $\xi$, the column density and
outflowing velocity for each component.  Tables~\ref{tab:n5548a} and
\ref{tab:n5548b} show a summary of the fitting parameters. The details will be
described in Seta et al. in prep.

\begin{table}[!htbp]
\caption{Summary of continuum parameters}\label{tab:n5548a}
\begin{center}
\begin{tabular}{lc}
\hline\hline
Parameter         & \\
\hline
power law: norm ($10^{52}$ ph/s/keV at 1 keV) & 0.56 $\pm$ 0.01\\
power law: $\Gamma$       & 1.72 $\pm$ 0.02\\ 
modified blackbody: area$\times$ density ($10^{33}$~cm$^{0.5}$) & 2.21 $\pm$ 0.16\\
modified blackbody:  $T$ (in eV)  & 125  $\pm$ 4\\
\hline
\end{tabular}
\end{center}
\end{table}

\begin{table}[!htbp]
\caption{Summary of absorption parameters}\label{tab:n5548b}
\begin{center}
\begin{tabular}{lcccc}
\hline\hline
 Comp. & {\rm log} $\xi$ & N$_{\rm H}$          & RMS velocity & Average velocity\\
       &                 & (10$^{21}$ cm$^{-2}$) & (km/s)       & (km/s)\\  
\hline
a & 1.14 $\pm$ 0.05         & 0.60 $\pm$ 0.07         & 95.3 $_{-17.8}^{+13.1}$ & $-$433.5 $_{-28.4}^{+24.3}$\\ 
b & 1.99 $_{-0.05}^{+0.03}$ & 0.86 $_{-0.15}^{+0.17}$ & 44.4 $_{-14.6}^{+26.1}$ & $-$276.9 $_{-47.7}^{+51.3}$\\ 
c & 2.34 $\pm$ 0.03         & 3.77 $_{-0.68}^{+0.80}$ & 33.3 $_{-6.1}^{+6.9}$   & $-$967.9 $_{-25.9}^{+26.6}$\\ 
d & 3.07 $\pm$ 0.04         & 7.19 $_{-1.56}^{+2.50}$ & 88.8 $_{-31.7}^{+37.1}$ & $-$479.6 $_{-62.1}^{+74.7}$\\ 
\hline
\end{tabular}
\end{center}
\end{table}

\begin{figure}[!htbp]
\begin{center}
\resizebox{0.43\hsize}{!}{\includegraphics[]{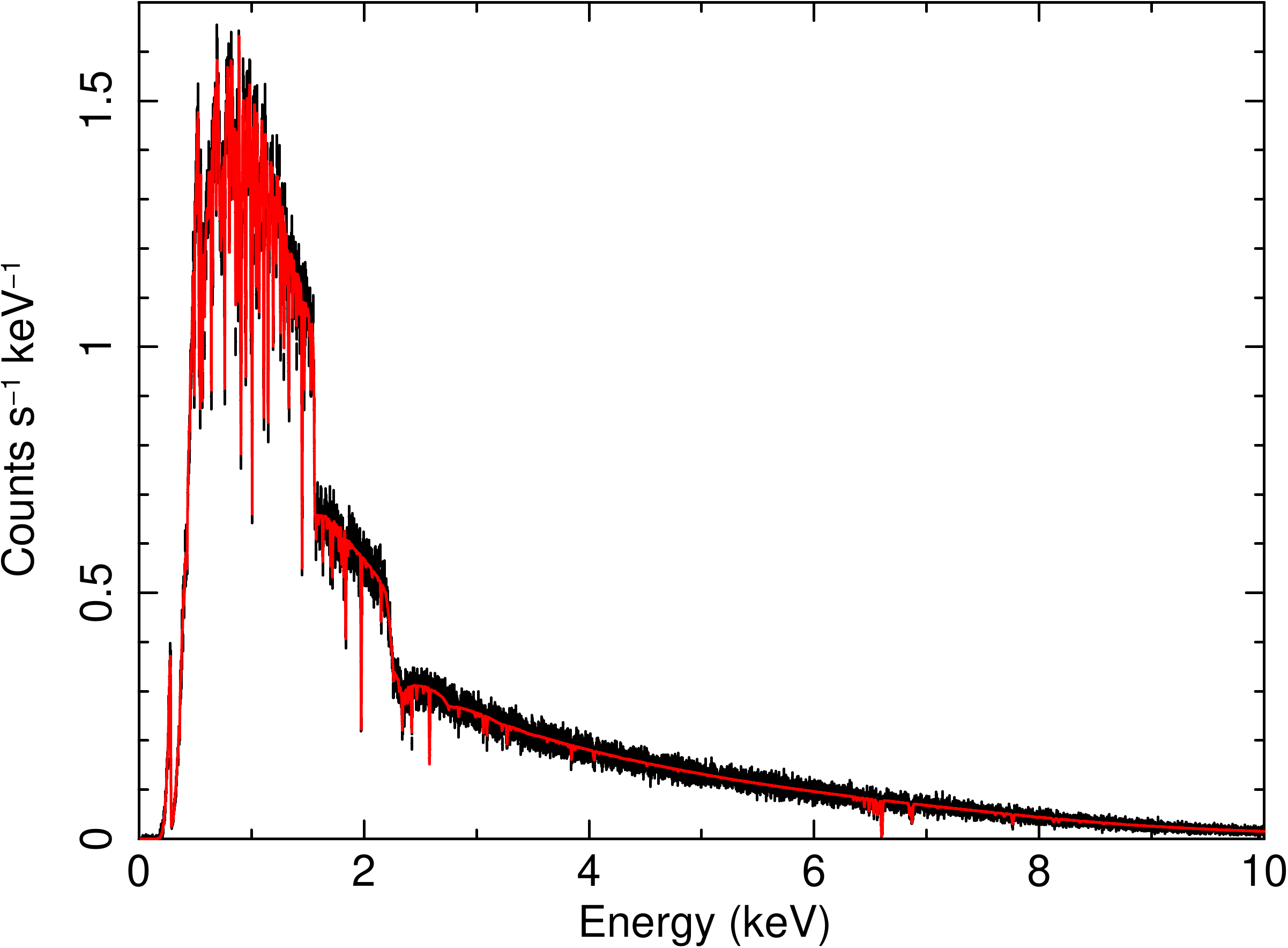}}
\resizebox{0.43\hsize}{!}{\includegraphics[]{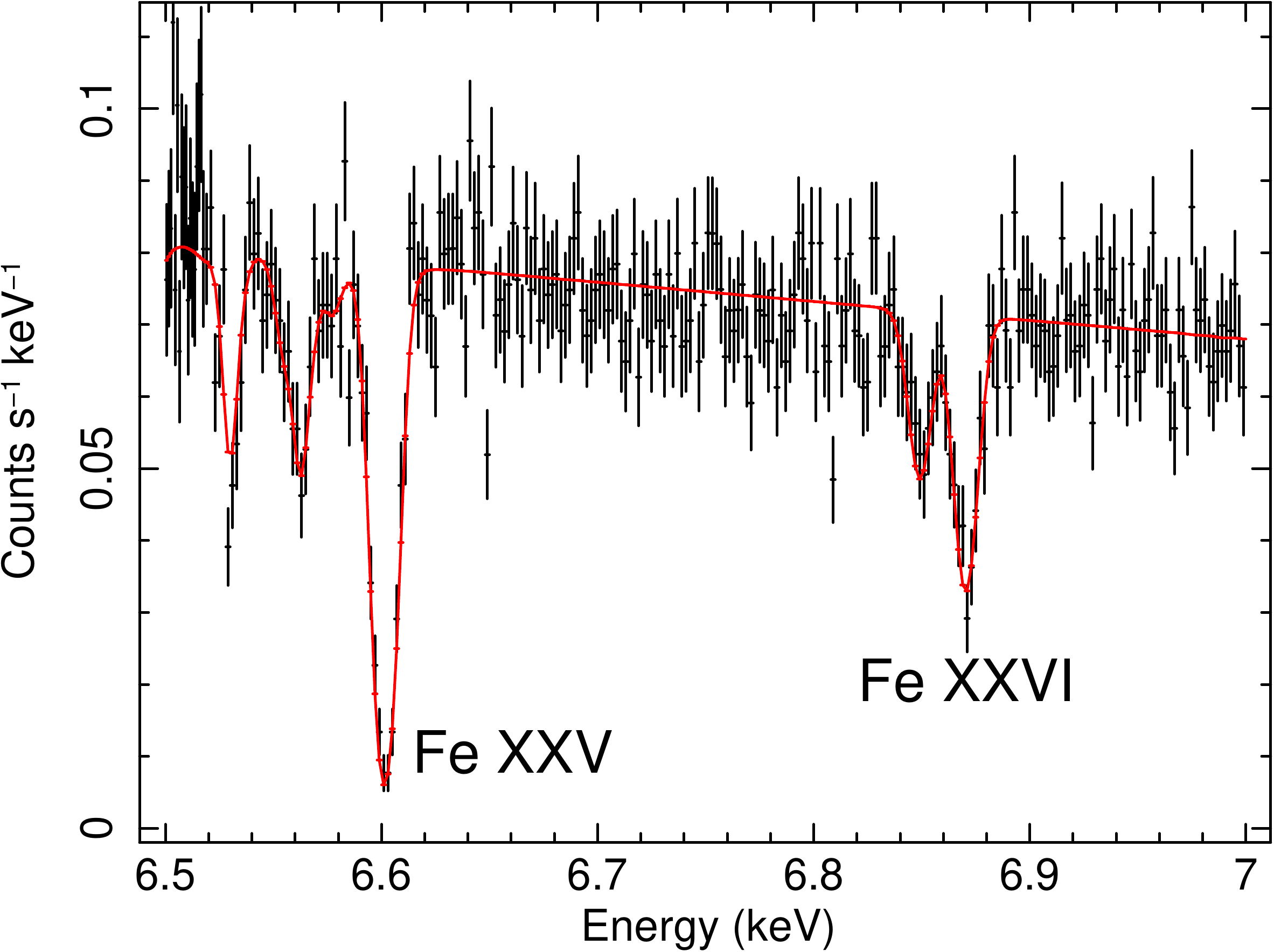}}
\end{center}
\caption{Simulated spectrum for NGC~5548 for 700~ks exposure time.}
\label{fig:ngc5548}
\end{figure}

We simulated the spectrum of NGC~5548 based on these parameters. We added two
components with Fe~XXV and Fe~XXVI columns. When the values of the Fe~XXV and
Fe~XXVI columns are about $1.5\times 10^{21}$~cm$^{-2}$, we found that we can
detect the Fe~XXV column with an integration time of 2~Ms at 3$\sigma$. On the
other hand, the Fe~XXVI column is undetectable. When the values of the Fe~XXV
and Fe~XXVI columns are ten times larger than this, we found that we can detect
the Fe~XXV and Fe~XXVI columns with an integration time of 700~ks at $3\sigma$.
The left panel of Figure~\ref{fig:ngc5548} shows the simulated spectrum of a
700~ks exposure time. The right panel of Figure~\ref{fig:ngc5548} shows a
close-up view of the Fe~XXV and Fe~XXVI absorption lines.

\subsubsection{NGC~3516}

\begin{figure}[!htbp]
\begin{center}
\includegraphics[width=0.65\hsize]{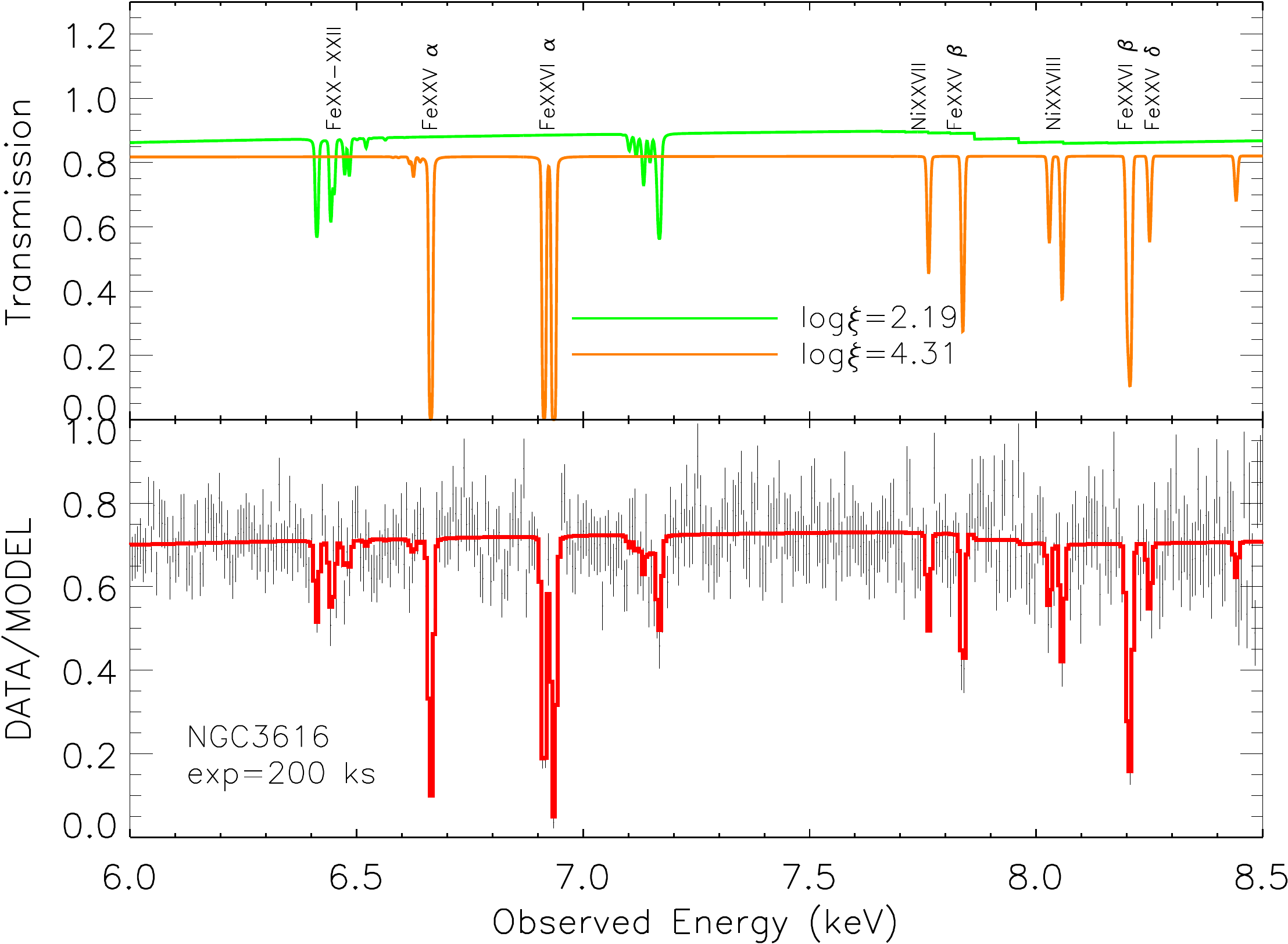}
\end{center}
\caption{Simulated spectrum for NGC~3516 for 200~ks exposure time.}
\label{fig:ngc3516}
\end{figure}

We simulated the spectrum of NGC~3516 based on the {\it XMM-Newton} EPIC
observation  published by \citep{turner08}.  The exposure time of 200~ks was
chosen to detect the ionic column density of Fe~XXV at least at 5$\sigma$
significance. Note the saturation of the 1s--2p lines (Fe~XXV and Fe~XXVI, see
Figure~\ref{fig:ngc3516}). The resulting best-fit column densities are $(4.8\pm
0.8)\times 10^{17}$~cm$^{-2}$ and $(2.6\pm 0.4)\times 10^{18}$~cm$^{-2}$ for
Fe~XXV and Fe~XXVI, respectively.

The model consists of 4 warm absorbers, but only two have a visible effect in
the iron region, see Figure~\ref{fig:ngc3516}. They are modelled with the
\textsl{xabs} model of SPEX. The components have hydrogen column densities (in
units of $10^{16}$~cm$^{-2}$) of 24, 5, 2020 and 2620, respectively, with
$\log\xi$-values of $-2.43$, 0.25, 2.19 and 4.31. Components 1 and 2 have zero
outflow velocity, while component 3 has an outflow velocity of
$-1600$~km\,s$^{-1}$ and component 4 of $-1000$~km\,s$^{-1}$. Component 3 has a
covering factor of 0.45, while all other components have covering factor unity.
The continuum is a simple power-law with photon index 1.9 and 2--10~keV flux of
$5.4\times 10^{-11}$ cgs.

Note that the source is highly variable in the 0.5--2 keV band (orders of
magnitude),  but not extremely variable in the medium band (2--10 keV, max
variation a factor of 5). We simulated a relatively high state, recorded in
2006. The parameters taken from {\it XMM-Newton} EPIC data analysis \citep{turner08}.

\subsubsection{H 0557-385}

\begin{figure}[!htbp]
\begin{center}
\resizebox{0.48\hsize}{!}{\includegraphics[angle=-90]{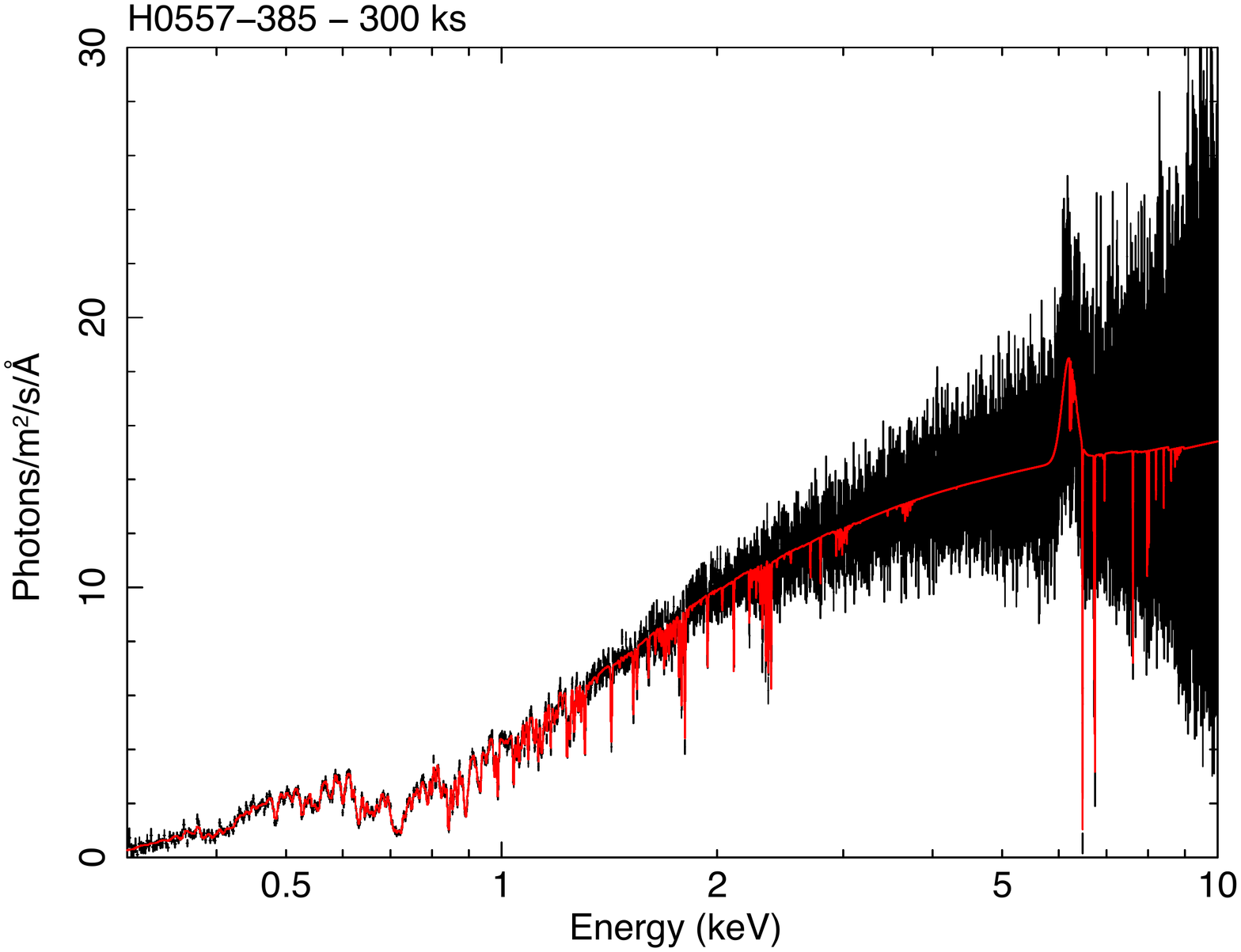}}
\,\,
\resizebox{0.48\hsize}{!}{\includegraphics[angle=-90]{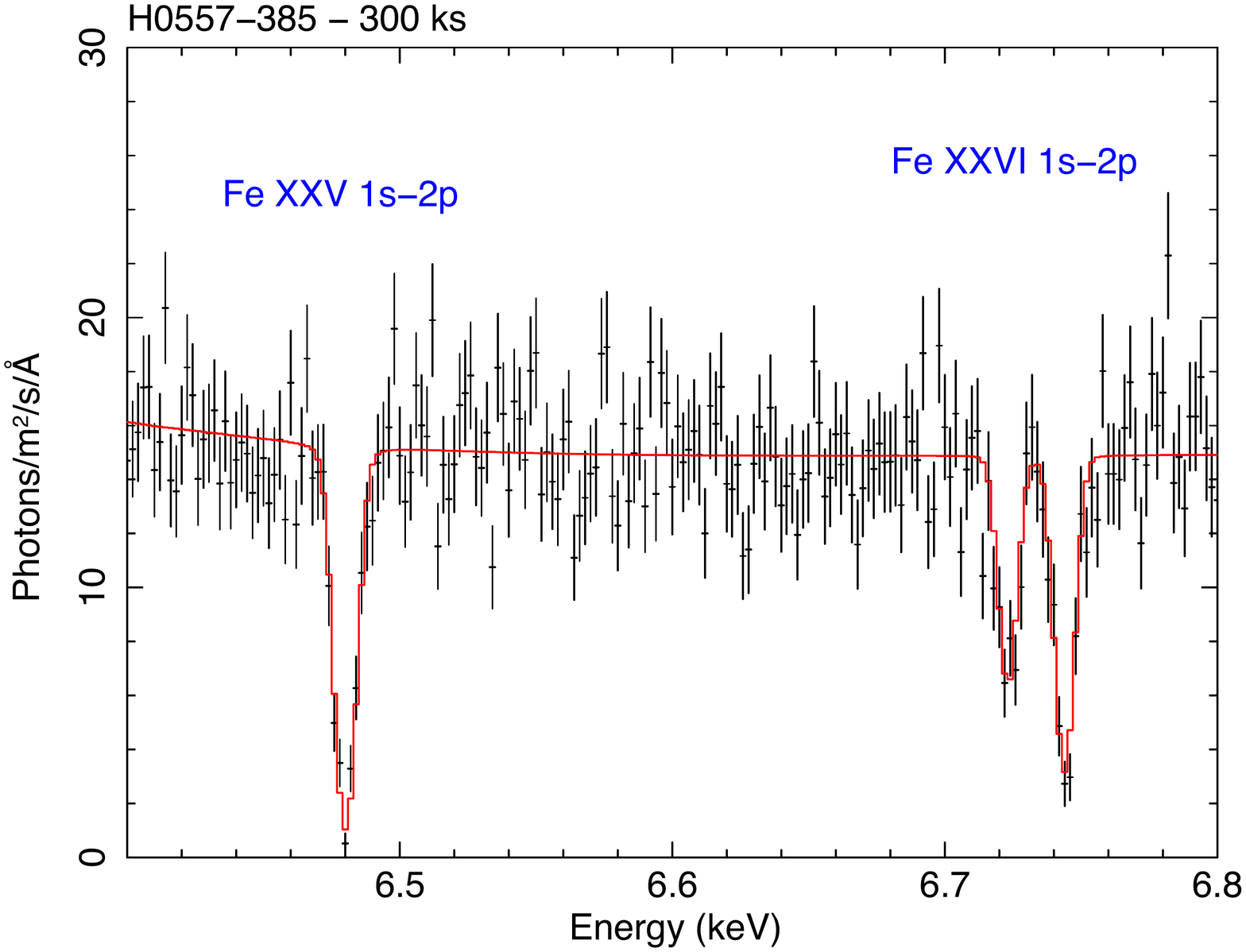}}
\end{center}
\caption{Simulated spectrum for H~0557-385 for 300~ks exposure time. See also
the remark at Figure~\ref{fig:ngc3783}.}
\label{fig:h0557}
\end{figure}

We simulated the spectrum of H~0557-385 based on the {\it XMM-Newton}
observation  published by \citet{ashton06}. The model includes two lower-$\xi$
components observed with {\it XMM-Newton}. We add two components with Fe~XXV
and Fe~XXVI columns of 4 and $5\times 10^{21}$~cm$^{-2}$, respectively, and
found that for an integration time of 200~ks we can measure the Fe~XXV and
Fe~XXVI columns with 20 and 38\% accuracy, respectively. To get Fe~XXVI to the
20\% level would require 500~ks exposure time. An exposure time of 300~ks
appears to be a good compromise and has been used for the simulation shown in
Figure~\ref{fig:h0557}. The optical depth of the strongest 1s--2p lines at line
center is 6.2 for Fe~XXV and 2.6 for Fe~XXVI.

We note here that H~0557-385 is usually in a bright state corresponding to our
simulated spectrum. However, it has been observed once in a ten times dimmer
state with very interesting Fe-K emission line properties \citep{longinotti09}
ascribed to intervening line-of-sight clouds.

\subsection{Beyond Feasibility}

Time variability of the outflow can be used to determine or constrain the
absolute location of the outflowing gas, by measuring delays of the ionisation
state with respect to changes in the ionising continuum. This delay is
essentially the recombination time scale, that scales inverse proportional to
the density. Most observations of outflows with {\it ASTRO-H} will likely be
long, 100~ks or more, which means that the total duration of the observation
will be a few days. Less massive AGN will show significant continuum variations
during the observation and therefore allow to constrain recombination time
scales at the day time scale. For more massive AGN, or for less massive AGN for
longer time scales, a second observation separated by weeks or half a year from
the first one will allow the same study but for longer time scales.

Further, SXS is most sensitive to the highest ionised gas. In order to derive
the full absorption measure distribution, simultaneous observations with the
{\it Chandra} or {\it XMM-Newton} gratings will allow for such a complete
picture. In this respect in particular the {\it Chandra} LETG and
{\it XMM-Newton} RGS are most useful, because they are sensitive to the
softest X-rays. The use of {\it XMM-Newton} has the additional benefit of
yielding simultaneously the UV continuum through data from the Optical Monitor.
This UV continuum is needed for accurate photoionisation calculations.

Finally, depending on the properties of the outflow, it may be possible to
measure the covering factor of the X-ray source by the outflow, therefore
providing a direct estimate of its size. This can be achieved by comparing the
intensity of two or more lines from the same ion, for instance the Ly$\alpha$
doublet of Fe~XXVI.

\section{Ultra Fast Outflows} 

\subsection{Background and Previous Studies}

The advent of the high throughput CCDs on board {\it XMM-Newton},
{\it Chandra} and later {\it Suzaku} brought also additional, somewhat
unexpected discoveries in the field of X-ray spectroscopy and in particular
regarding AGN outflows.

For instance, \citet{chartas02,chartas03} reported the first evidences for BAL
quasar type of outflows in the {\it Chandra} and {\it XMM-Newton} X-ray
spectra of the gravitationally lensed, high redshift ($z \ge 2$) quasars
APM~08279$+$5255 and PG~1115$+$080 through the detection of absorption lines at
rest-frame energies E$>$7~keV. Meanwhile, \citet{pounds03} reported the
detection with {\it XMM-Newton} of similar absorption lines also in the local
($z=0.08$) quasar PG~1211$+$143, which then became the archetype for such type
of outflows. If identified with Fe~XXV/XXVI absorption lines, these features
suggest the presence of previously unknown highly ionized zones of circumnuclear
absorbing material with large column densities and blue-shifted velocities
reaching mildly-relativistic values of $\sim$0.1--0.3c, which are possibly
directly connected with accretion disk winds/outflows.

Indeed, detailed theoretical studies of accretion disk winds in AGNs show that
the inner parts of the outflow can be highly ionized by the intense nuclear
radiation and, in principle, can easily reach mildly-relativistic velocities
\citep[e.g.][]{king03,ohsuga09,king10,fukumura10}. Therefore, when the line of
sight to the observer intercepts the flow, considerable absorption features from
highly ionized species can be imprinted in the X-ray spectrum in the form of
blue-shifted Fe XXV/XXVI K-shell absorption lines at rest-frame energies greater
than $\sim$7~keV.

Following these first seminal cases, several papers reporting the detection of
similar blue-shifted absorption lines in a number of low-redshift Seyfert
galaxies and quasars appeared in the literature as well, although the claimed
features often had only a limited statistical significance of
$\simeq$3--4$\sigma$. In fact, an important drawback regarding these features is
due to the fact that for low-redshift sources (representing the majority of the
cases) they appear at energies E$>$7~keV, a region of the spectrum where the
instrumental resolution and signal-to-noise of the present X-ray instruments are
very limited. Moreover, due to its high ionization, this material is hard to
observe in soft X-ray spectra because all the elements lighter than Fe are
almost completely stripped of their electrons and this may be one of the reasons
why it was not reported previously.

Nevertheless, the first systematic and comprehensive search for blue-shifted Fe
K absorption lines at E$\ge$7~keV in a large sample of 42 local ($z \le 0.1$)
Seyferts observed with {\it XMM-Newton} was performed by \citet{tombesi10a}.
They estimated the absorption line parameters carrying out a uniform modeling of
all datasets and calculated the detection significance making use of both the
F-test and extensive Monte Carlo simulations. In order to mark an initial
distinction with the more typical warm absorbers, which usually show maximum
velocities lower than a few 1\,000~km\,s$^{-1}$, \citet{tombesi10a} also defined
ultra-fast outflows (UFOs) as those highly ionized absorbers detected
essentially through Fe~XXV and/or Fe~XXVI lines with a blue-shifted velocity
higher than 10\,000~km\,s$^{-1}$ ($\simeq$0.03c). The majority of the detected
lines are consistent with UFOs and their observed fraction is $\sim$30--40\%,
indicating that UFOs could possibly represent a common phenomenon taking place
in these radio-quiet AGNs. Although hampered by the limited CCD sensitivity, the
data indicate possible variability of some of the line equivalent widths and
velocity shifts among different observations of the same sources, even on
time-scales shorter than days, possibly indicating compact absorbers.
Noteworthy, these results have been confirmed also by an independent analysis of
a large sample of AGNs performed by \citet{gofford12} using broad-band
{\it Suzaku} data.

Despite the significant uncertainties due to the limited CCD sensitivity and
energy resolution, photo-ionization modeling of the blue-shifted Fe K absorption
lines associated with UFOs has been performed using codes such as \emph{Xstar}
\citep[e.g.,][]{tombesi11,gofford12}. As anticipated, the outflow velocity
distribution is found to be mildly-relativistic, in the range $\simeq$0.03--0.3c
($\simeq$10\,000--100\,000~km~s$^{-1}$), with mean value of $\simeq$0.1c. The
ionization is in the range log$\xi \simeq$3--6 and the column densities are in
the interval $N_{\rm H}\simeq 10^{22}$--$10^{24}$~cm$^{-2}$. Some of these
models will be used as templates to carry out the realistic {\it ASTRO-H}
calorimeter simulations reported in the subsequent sections. 

Several papers have been published attempting to estimate the location and
energetics of these UFOs. However, due to the large uncertainties, they have
been able to place only order of magnitude lower/upper limits. For instance,
\citet{tombesi12} performed a uniform investigation of these parameters in a
large sample of Seyferts galaxies. On average, the UFOs seem to be observed at
sub-pc distances from the central supermassive black hole, corresponding to
$\simeq$$10^2$--$10^4$ Schwarzschild radii. Their mass outflow rates are
probably in the range between $\simeq$0.01--1~M$_{\odot}$yr$^{-1}$ and, given
their high velocity, their associated mechanical power can easily reach very
high values of $\dot{E}_K$$\simeq$$10^{42}$--$10^{44}$~erg~s$^{-1}$, which might
be even comparable to that of some of the jets in radio galaxies. In particular,
these values are systematically larger than $\sim$0.5\% of the bolometric
luminosity, which is the minimum ratio required by numerical simulations of
feedback induced by AGN winds/outflows \citep[e.g.,][]{hopkins10}.

Therefore, overall, these results point to the conclusion that such UFOs might
indeed have the possibility to bring outward a significant amount of mass and
energy from the central regions of AGNs, which can then have an important
influence on the surrounding environment \citep[e.g.,][]{cappi06}. Feedback from
AGNs is expected to have a significant role in the evolution of the host
galaxies, such as the enrichment of the ISM or the quenching of star formation,
and could also contribute to the establishment of some fundamental relations,
such as the $M_{BH}$--$\sigma$. The ejection of a significant amount of mass
from the central regions of AGNs could also inhibit the growth of supermassive
black holes, potentially affecting their evolution. Therefore, the UFOs might
well represent the long sought-after agent mediating the ``quasar mode'' galaxy
feedback.

The UFOs, and some of the other more ``normal'' AGN outflows
\citep[e.g.,][]{crenshaw12}, might actually provide a feedback impact comparable
or even greater than that from jets. In fact, the UFOs are likely more massive
than jets. They are mildly-relativistic and have somewhat wide angles, therefore
possibly exerting a higher impact on the surrounding host galaxy environment
compared to the highly collimated relativistic jets, which might actually drill
out of the galaxy and have a dominant effect only in the outside. UFOs are
energetic, with a mechanical power possibly comparable to that of jets.
Moreover, evidences for UFOs have been found in at least $\sim$30--40\% of the
local radio-quiet AGNs and may possibly have a more widespread feedback
influence with respect to the less common radio-loud sources with powerful
jets. 

In this respect, it is important to note that evidences for UFOs have been
reported in some radio-loud AGNs as well \citep{tombesi10b,gofford12} and
therefore their feedback effect might actually be concomitant with that from
jets. This opens also a new line of investigation that can help to shed light on
the intricate connection between the accretion disk and the formation of jets
and outflows, similarly to what has been reported for the radio galaxies 3C~111
and 3C~120 \citep{chatterjee09,chatterjee11} and the micro-quasar
GRS~19151$+$105 \citep{neilsen09}. Comparing the mechanical power of winds and
jets from stellar mass to supermassive black holes, \citet{king13} recently
suggested that the characteristics of UFOs tend to resemble those of jets more
than winds, indicating that we may be observing a transition regime in which
winds become jets. This, as well as other evidences, may open the unexpected
possibility to effectively study jet related events through X-ray absorption
line spectroscopy. 

We note that some of the AGNs in which evidences for UFOs have been reported
also simultaneously show the more ``normal'' outflows in the soft X-rays. This
rises the issue of what is, if any, the connection between these two type of
ionized absorbers. One of the first attempts to find an answer to this question
was reported very recently by \citet{tombesi13}, who performed a comparison of
the characteristics of the UFOs and the warm absorbers in a sample Seyfert 1
galaxies observed with {\it XMM-Newton} and {\it Chandra}. For the first
time, they claim the existence of correlations between the different parameters,
indicating that the closer the absorber is to the central black hole, the higher
the ionization, column density, outflow velocity and consequently the mechanical
power. This suggests that these absorbers could actually be unified as parts of
a single, large-scale stratified outflow observed at different locations along
the line of sight, from the vicinity of the central supermassive black hole up
to the outskirts of the host galaxy. The actual acceleration mechanism(s) of
these outflows are still unknown and this represents one of the main challenges
posed to both observation and theory. For instance, \citet{tombesi13} suggests
that the observations are consistent with a combination of radiation pressure
through Compton scattering \citep[e.g.,][]{king03} and/or MHD processes
\citep[e.g.,][]{fukumura10}, the latter being in agreement with the wind-jet
connection suggested by \citet{king13}.

\subsection{Prospects \& Strategy}

In the previous section we described a series of amazing scientific implications
and the broad impact on the physics of AGNs and galaxy evolution brought by the
discovery of UFOs. However, it is important to note that there is a fundamental
drawback in the study of UFOs through blue-shifted Fe~XXV-XXVI absorption lines
at E$>$7~keV with current CCD instruments. Often, just one absorption line is
detectable and the great majority of the detections are weak, in the range
between $\sim$3--4$\sigma$. Since the initial claims of detections, this raised
strong debates and criticisms in the community and, despite the improvements in
recent years, the issue has not been fully solved yet and there are even
scepticisms about the actual existence of these UFOs. In the next sections we
will show that the unprecedented energy resolution and sensitivity in the
crucial E$=$4--10~keV band provided by the {\it ASTRO-H} calorimeter will
allow a fundamental breakthrough in this field and the definitive solution of
this problem.   

One major point will be not just the increased detection significance of each
single weak absorption line, but the possibility to simultaneously detect more
than one line from the same photo-ionized absorber. From a previous
{\it XMM-Newton} sample study of blue-shifted Fe~XXV-XXVI absorption lines,
\citet{tombesi11} reported that in 19 cases the lines were consistent with UFOs
and for each of these only a single significant line was detected. Using the
photo-ionization code {\it Xstar}, they identified 9 lines with Fe~XXVI, 4
with Fe~XXV and for 6 of them the identification was degenerate (at the 90\%
level) between Fe~XXV or Fe~XXVI. Instead, \citet{gofford12} using
{\it Suzaku} reported lines consistent with UFOs in 12 cases, 6 of which were
single and identified with Fe~XXVI, 3 single but degenerate between Fe~XXV or
Fe~XXVI, 2 blended and 1 double identified with both Fe~XXV and Fe~XXVI
together. With {\it ASTRO-H}, we expect the number of detected double lines
to significantly increase, drastically improving the significance of the
observed UFOs. Additional, higher order transitions of Fe~XXV-XXVI and
absorption lines from lighter elements as well (such as S, Si, Ca, Ar) could be
possibly detected by {\it ASTRO-H}.

For instance, if we detect two lines each with null hypothesis significances
$p_1$ and $p_2$ of $\sim$3$\sigma$ (corresponding to $\sim$99.9\% chance
probability), their combined detection level $P\simeq p_1 \times p_2$
effectively increases to $\sim$5$\sigma$ ($\sim$99.9999\% chance probability).
Thus, the simultaneous detection of multiple lines from Fe~XXV and/or Fe~XXVI
will tremendously boost the detection significance of the associated UFOs and
will drastically lower the minimum sensible column densities (even down to $N_H
\sim 10^{20}$~cm$^{-2}$). Considering the full E$\simeq$0.5--10~keV band,
{\it ASTRO-H} will allow also to simultaneously detect different absorption
components, from the soft X-rays to the hard X-rays, and dramatically expand the
observed range in ionization, column density and outflow velocity. The expected
detection of several lines in the Fe K band will potentially provide a
breakthrough comparable to that brought to the warm absorber studies by the
advent of the grating spectrometers on board {\it Chandra} and
{\it XMM-Newton}. Therefore, it is likely that with {\it ASTRO-H} the UFO
could actually become the new ``normal'' type of outflows.

Only a handful of the absorption lines detected with the CCDs have been
resolved. In general, their velocity broadening ($\sigma_v$) is expected to be
in the range between $\sim$1\,000~km/s up to $\sim$5\,000~km/s. The calorimeter
will allow to resolve all these lines and to even study the profile for the
broadest ones, drastically decreasing the uncertainties on the column density
and possibly helping to test the different profiles expected from radiation or
MHD acceleration mechanisms \citep[e.g.,][]{fukumura10}. Due to the limited
energy resolution of the present CCDs, in many cases the lines appear also
blended together and their identification can be degenerate between Fe~XXV
and/or Fe~XXVI. The detection of more lines with the calorimeter will allow to
better discriminate between different ionization levels.

Finally, we note that previous CCD observations suggested that the blue-shifted
Fe XXV-XXVI absorption lines may show variability in both EW and velocity shift
even on time scales shorter than days, in accordance with the idea that the UFOs
are ejected from the inner accretion disk close to the black hole and that they
represent the most dynamically important outflow component. Their variability
could be due to a response to changes in the source luminosity or could be
intrinsic, possibly depending also on their unknown ejection duty cycle. In
general, studies of large samples find a detection frequency of $\sim$30\%
\citep{tombesi10a,gofford12}. Therefore, in order to take into account the
possibility of detection/non-detection due to their intermittent nature, our
preferred strategy should be on observing more sources or one source multiple
times. For instance, from the observation of 10 sources we expect to detect UFOs
in at least $\sim$3 of them or, equivalently, we expect at least one secure
detection out of $\sim$3 observations. This point is very important because it
represents another of the main criticisms on the actual existence of UFOs, which
says that the ``spurious'' detection of weak lines in different observations could
just indicate random noise. 

In the next sections we will better quantify these claims, we will derive some
exposure time estimates for a sample of possible candidate sources and we will
show some realistic simulated spectra.

\subsection{Targets \& Feasibility}

We selected the targets from the local ($z < 0.1$) AGNs with detected UFOs
reported in \citet{tombesi10a,tombesi10b} and \citet{gofford12}. The equivalent
widths of the Fe~XXV and/or Fe~XXVI absorption lines relative to the UFOs are in
the range from $\simeq$15~eV up to $\simeq$100~eV, with mean and median values
of $\simeq$50~eV and $\simeq$30~eV, respectively. The average blue-shifted
velocity of these lines is $\simeq$0.1c, which corresponds to an observed energy
of E$\simeq$8~keV. Therefore, we consider the most conservative case assuming
only one line at 8~keV (where the effective area is $\simeq$200~cm$^{-2}$) with
the minimum observed equivalent width of EW$=$15~eV and for each source we
consider the minimum 4--10~keV flux level among the relative observations. Then,
we use equation~\ref{eqn:strongline} to calculate the minimum exposure time
required to reach a detection confidence level of 5$\sigma$. The approximate
exposure times for all the candidate sources were calculated. Due to the fast
decrease in effective area, to detect the same line at $\sim$9~keV,
corresponding to a velocity shift of $\sim$0.2c, it would require an increase of
$\sim$60\% in exposure. We find that for sources with 4--10~keV flux $\ge
1\times 10^{-11}$~erg~s$^{-1}$~cm$^{-2}$ it will be even possible to also study
very short time-scale variability on $\sim$20~ks time-scales. In the next
sections we show a few dedicated simulations for {\it ASTRO-H} relative to
some of the most promising sources.

\subsubsection{Mrk~509}

\begin{figure}[!htbp]
\begin{center}
\resizebox{0.60\hsize}{!}{\includegraphics[angle=-90]{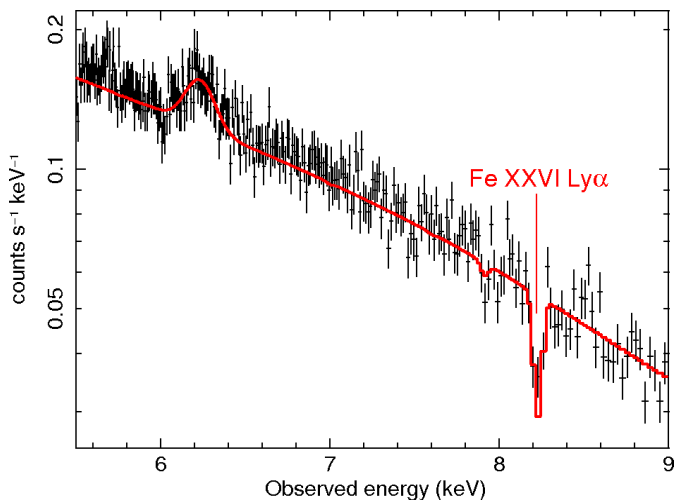}}
\end{center}
\caption{ Simulated 100~ks {\it ASTRO-H}
calorimeter spectrum of Mrk~509 using the {\it XMM-Newton} observation of April 2006
as baseline. The highly ionized UFO has a velocity $\simeq$0.2c. The dataset has
been grouped to a S/N of 10 for each energy bin.}
\label{fig:mrk509}
\end{figure}

We performed a 100~ks realistic {\it ASTRO-H} calorimeter simulation of the
Seyfert 1 galaxy Mrk~509, which has been claimed to show UFOs in some of its
{\it XMM-Newton} observations. We assume the same spectrum as of the
{\it XMM-Newton} observation of April 2006 in which the detection of a weak
blue-shifted absorption line with EW$\simeq$20~eV at the rest-frame energy of
$\simeq$8.5~keV was reported \citep{cappi09,tombesi10a}. Due to the limited CCD
resolution and sensitivity, it was not possible to discriminate between Fe~XXV
or Fe~XXVI and a fit with an \emph{Xstar} table with turbulent velocity of
1,000~km\,s$^{-1}$ gave two relative solutions with average outflow velocity of
$\sim$0.2c. We focus in the E$=$4--10~keV band and assume the Fe~XXVI model of
\citet{tombesi11}. The simulated spectrum is shown in Figure~\ref{fig:mrk509}.
Imposing a fit with the Fe~XXV model, we find that it is rejected with a
$\Delta\chi^2 \sim 33$, which corresponds to a $\sim 5 \sigma$ level. Therefore,
{\it ASTRO-H} will indeed be able to clearly discriminate between the two
possible solutions. Instead, fitting with the Fe~XXVI model, we find an
improvement of $\Delta\chi^2 \sim 80$, indicating that the UFO will be
detectable at $\gg$5$\sigma$ (compared only to a $\Delta\chi^2 \sim 15$ of
{\it XMM-Newton}).

\subsubsection{3C~111}

We performed a 100~ks realistic {\it ASTRO-H} calorimeter simulation of the
broad-line radio galaxy 3C~111 based on the 100~ks {\it Suzaku} XIS spectrum
taken in 2008. We focused only in the E$=$4--10~keV band. Blue-shifted Fe K
absorption lines were detected in this spectrum by \citet{tombesi10b} and
\citet{gofford12}, indicating that this is one of the few radio-loud AGN showing
UFOs to date. Interestingly, this source shows also a powerful, time variable
relativistic radio jet and was detected in the $\gamma$-rays with the Fermi
satellite. Therefore, the study of its UFO has important implications for the
investigation of the possible connection between the disk, the jet and outflows
and their feedback on the host galaxy. 

\begin{figure}[!htbp]
\begin{center}
\resizebox{0.48\hsize}{!}{\includegraphics[angle=0]{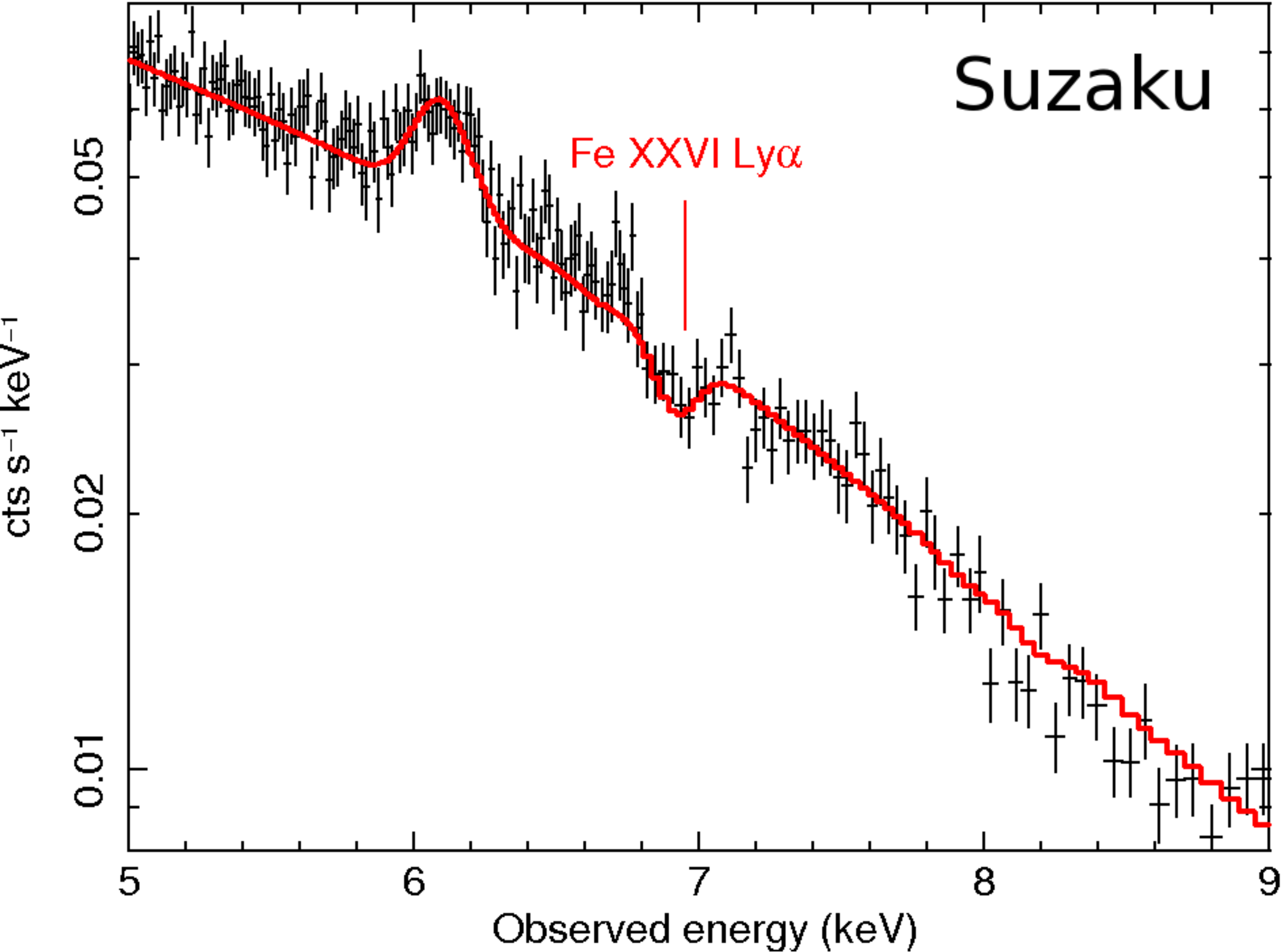}}
\,\,
\resizebox{0.48\hsize}{!}{\includegraphics[angle=0]{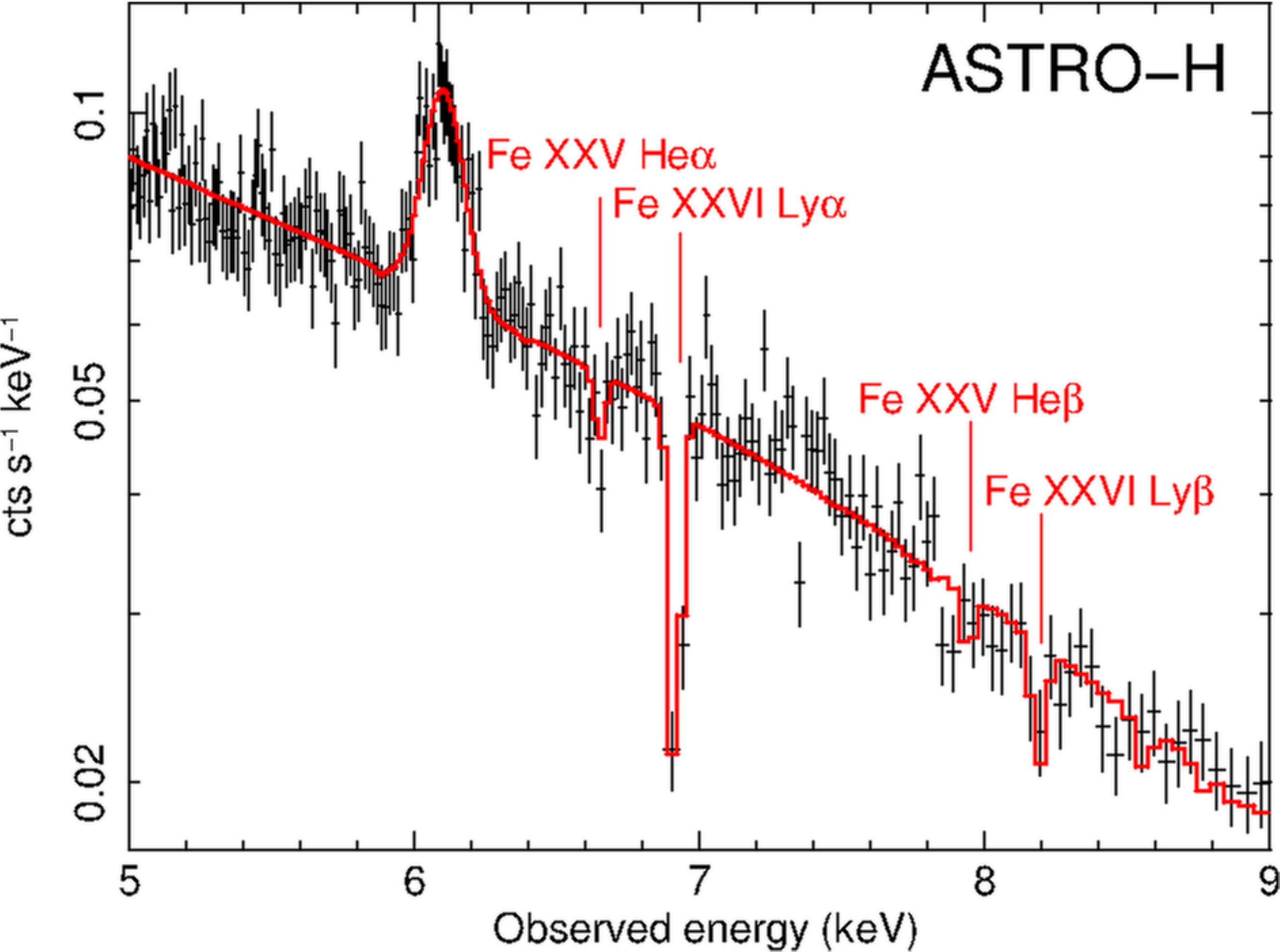}}
\end{center}
\caption{ Spectrum of the radio galaxy 3C~111
showing a UFO with velocity $\simeq$0.04c. \emph{Left panel:} real 100~ks
combined {\it Suzaku} XIS~0$+$XIS~3 spectrum obtained in 2008. \emph{Right
panel:} simulated 100~ks {\it ASTRO-H} calorimeter spectrum. Both datasets have been
grouped to a S/N of 10 for each energy bin.}
\label{fig:3c111}
\end{figure}

We assume the best-fit model of \citet{tombesi10b}, which is composed of a
Galactic absorbed power-law continuum with $\Gamma \simeq 1.5$, a neutral Fe
K$\alpha$ emission line and an \emph{Xstar} component with turbulent velocity of
1\,000~km\,s$^{-1}$, column density $N_H\simeq 10^{23}$~cm$^{-2}$, high
ionization log$\xi$$\simeq$5 and outflow velocity of $v_{out}\simeq 0.04$c. We
note that during this observation the source was caught in an historical flux
minimum of $\sim 1\times 10^{-11}$~erg~s$^{-1}$~cm$^{-2}$. From
Figure~\ref{fig:3c111} we can clearly note the huge improvement in the detection
of the blue-shifted absorption lines between the 100~ks combined {\it Suzaku}
XIS~0$+$XIS~3 spectrum on the left and the simulated 100~ks {\it ASTRO-H}
calorimeter on the right. Many more absorption lines are clearly observable with
the calorimeter. The fit improvement given by the addition of the \emph{Xstar}
absorption component in the calorimeter spectrum is enormous, $\Delta\chi^2 =
280$ for three free parameters, instead of the mere $\Delta\chi^2 \simeq 22$ of
the {\it Suzaku} observation. This would allow to clear any doubts about the
real detection of the UFO, which will be detectable at a $\sim 20 \sigma$
significance level. The estimates of the absorber parameters will be
significantly improved as well, for instance the column density will be
estimated with $\sim$15\% errors (presently only lower limit with
{\it Suzaku}), the ionization with 2\%  and the velocity with $<$1\% (both
$\sim$10\% with {\it Suzaku}). For the most intense absorption line, the
Fe~XXVI Ly$\alpha$, the centroid energy will be constrained with 0.05\% errors
(10 times better than {\it Suzaku}), the line will be resolved with $<$10\%
errors (presently not resolved {\it Suzaku}), allowing to estimate the
velocity broadening, and the EW will be estimated with 5\% error.

\subsubsection{PG~1211+143}

PG~1211+143 ($z = 0.0809$) is the quasar in which \citet{pounds03} first
reported the detection of highly ionized blueshifted iron absorption features. A
subsequent analysis with detailed photoionization modeling by Tombesi et al.
(2011), these features are well reproduced by an highly ionized ($\log\xi=
2.87$) outflowing ($\mathrm{v}_{\rm out}= 0.15$c) gas along the line of sight
(with column density of about 10$^{23}$~cm$^{-2}$). 

Here, we present the results of the SXS calorimeter simulation of PG~1211+143
based on the {\it XMM-Newton} observation in 2001. We used the redistribution
matrix function (RMF) with 5~eV resolution, which is the current best-estimated
version (ah\_sxs\_5ev\_basefi lt\_20100712.rmf), and the ancillary response file
(ARF) for the point source (sxt-s\_120210\_ts02um\_of\_intallpxl.arf). The model
assumed is based on \citet{tombesi11}. Errors are quoted at 68\% confidence, for
one parameter of interest.

\begin{figure}[!htbp]
\begin{center}
\resizebox{0.5\hsize}{!}{\includegraphics[]{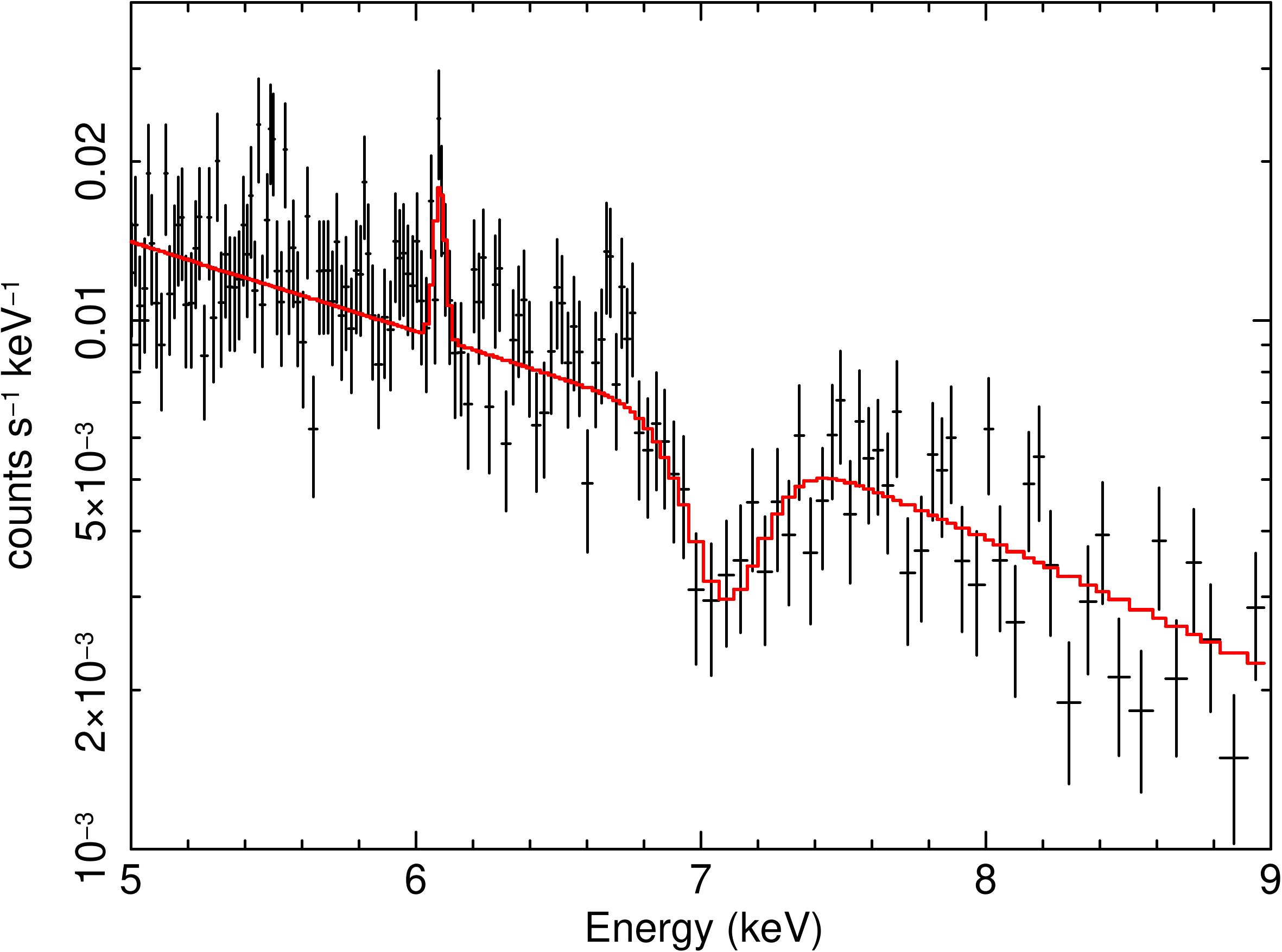}}
\end{center}
\caption{Simulated spectrum of PG~1211+143 for 100 ks
  observation. The absorption feature around 7 keV is due to Fe XXV.}
\label{fig:pg1211_sim_spec}
\end{figure}

Figure~\ref{fig:pg1211_sim_spec} shows the simulated spectrum in the 5--9 keV
energy band for 100~ks exposure. The broad and deep trough is clearly seen
around 7 keV. In order to estimate the significance of the absorption feature,
we applied a single power-law and a negative Gaussian model to the data. The
line center energy (source rest frame) and the broadening are determined to be
$E=7.65\pm 0.03$~keV and $\sigma = 156^{+33}_{-23}$~eV, respectively. The
equivalent width derived is $-$186$^{+58}_{-67}$~eV. In order to investigate the
absorption feature in more detail, we applied a power-law absorbed by an ionized
absorber calculated via XSTAR photoionization code (Kallman \& Bautista 2001).
We assumed the turbulent velocity to be 5\,000 km\,s$^{-1}$. The inclusion of
the XSTAR model improves the fit significantly ($\Delta \chi ^{2} = 52$ for
three additional parameters). We are able to strictly constrain the outflow
velocity $v_{\rm out}$ to be 0.154$^{+0.001}_{-0.002}$c. The column density and
ionization parameter are determined to be $N_{\rm H}= 1.0^{+0.3}_{-0.2} \times
10^{23}$~cm$^{-2}$ and $\log\xi= 2.9^{+0.2}_{-0.3}$, respectively.  

\section{Comparison with numerical models}

\subsection{Background and Previous Studies}

Most theoretical modelling of AGN winds as it relates to X-ray properties has
been directed toward the question of how the winds are created and what their
structure is. This has implications for understanding the effects of winds on
the UV region of the spectrum, and therefore on larger questions about AGN:
their mass budget, their black hole masses, and their evolutionary state. In
principle, construction of a model for testing against X-ray data should consist
of: (i) begin with a physical model of an AGN, i.e. the dynamics of the gas
affecting the observed spectrum, followed by (ii) a calculation or a realization
of the dynamical model, and then performing (iii) a realistic radiative transfer
calculation incorporating as many of the relevant physical process and then
would (iv) demonstrate that the the model could produce all of the ``important''
aspects of high S/N spectra of a group of AGN. 

So far, however, most of the attempts to characterize wind features have
involved simple applications of phenomenological models contained within XSPEC
or SPEX. These models are iteratively fitted to observed data in order to yield
values for a small number of free parameters. The models are designed for
flexibility, i.e. to be applicable to a wide range of possible geometries,
densities, elemental abundances, etc. A goal is to establish conditions in warm
absorbers, for example to estimate the amount of reflection from the disk, or to
determine ionization conditions. These models typically incorporate substantial
detail in the synthesis of the spectrum, in order to provide the maximum realism
if their assumptions are correct. For example, models for the X-ray spectrum
radiated by a plasma in coronal equilibrium, or models for the X-ray opacity of
a photoionized plasma, typically include potential contributions from $\sim
10^5$ lines or edges from $\sim 10^2 - 10^3$ ions. On the other hand, they
ignore or simplify fundamental questions associated with the geometry and
dynamics of the gas. Therefore the failure of such a model to provide a
satisfactory or plausible fit is a likely indicator of the breakdown of one or
more of the simplifying assumptions. 

The lack of detailed modelling reflects both the absence of a consensus on the
geometry of AGN and also the limited number of high S/N X-ray spectra of AGN. 
However, some progress toward the goal of a consistent physical model for AGN
outflows is being made. For, example, \citet{schurch07} and more recently by
\citet{saez11} treat the wind essentially as a 1-d outflow, which is reasonable
if one is only interested in the shape of the absorption. They define the wind
in terms of a series of shells of varying velocity, with distribution based on
physically consistent dynamical models. They calculate the state of the gas in
and the transmitted spectrum in each shell sequentially using XSTAR or CLOUDY,
modifying the input spectrum in the next shell to allow for changes in velocity,
and in this manner create spectra of warm absorbers. \citet{schurch07} find they
can produce spectra that fit observations of warm absorbers in this manner;
\citet{saez11} find they can reproduce the main characteristics of the ultrafast
outflow in APM~08279+5255.

\citet{sim08, sim10}, by contrast, start with a two-dimensional bi-conical
flow.  They adopt a kinematic, or parametrized, description of the wind defined
in terms of total mass loss, opening-angle, etc., calculate the ionization
structure of this wind, and then calculate the X-ray spectrum that would result.
They find that with the appropriate choice of parameters, which include mass
loss rates comparable to the Eddington accretion rate, that they can reproduce
with some precision the shape of the absorption and emission spectrum of Mrk 766
\citep{sim08} and PKS\,1211-143 \citep{sim10}. Interestingly, these spectral
simulations show a redward wing on Fe K$\alpha$, without including
gravitationally-redshifted fluorescence from from a disk. Instead, the redward
wing of Fe K$\alpha$ arises directly from the scattering from the receding
portion of the wind. \citet{tatum12} have used XSPEC and a small grid of models
created with this Monte Carlo radiative transfer code to model the spectra of a
small sample of Seyfert galaxies observed with {\it Suzaku}, suggesting that
a large fraction of the X-ray flux in these AGN is processed by the wind.
Whether these types of wind are like those that are found in AGN remains
controversial \citep[see, e.g][for a rejoinder on the physical underpinnings of
these types of winds]{reynolds12}. However, similar profiles are produced
directly from \cite{kurosawa09} 3-d hydrodynamical simulations \citep{sim12}.

\cite{sim08, sim10} envision the wind as arising from the accretion disk, at a
distance of $10^{14}$~{cm}, however an alternative view is that the wind
associated with the warm absorber arises from the region of the torus, at a
distance of order $10^{18}$~{cm} from the BH, the same torus that is believed to
obscure the broad line region in Seyfert II galaxies \citep{krolik95, krolik01}.
\cite{doronitsyn12}, for example, have recently calculated 3-d hydrodynamical
models showing that a bi-conical flow is generated by the intense flux of X-rays
at the inner edge of the torus showing that these flows can generate absorption
that resembles that observed from warm absorbers in Seyfert I galaxies. Typical
mass loss rates of range from about 10$^{-2}$ to 10$^{-1}$ {M$_\odot$\,yr$^{-1}$
and velocities of the outflowing gas are typically 100-1000 km s$^{-1}$. While
detailed fits to existing X-ray spectra of systems have not been attempted (no
doubt in part because the outflow is not steady), this is just a matter of time.

\subsection{Prospects \& Strategy}

The most widely applied models are based on extremely simple assumptions about 
the more fundamental physical quantities which affect the observations. For
example, time-steady equilibrium is assumed in the most widely  applied models,
and a single value of the most important free parameters (eg. temperature,
ionization balance) is specified rather than a distribution. The failure of such
a model to provide a satisfactory or plausible fit is therefore a likely
indicator of the breakdown of one or more of these assumptions. In this sense,
many or all {\it ASTRO-H} observations  will include the testing of numerical
models.  This type of model  testing is not unique to {\it ASTRO-H}, though
it is likely that  {\it ASTRO-H} will provide new tests or results of this
type owing to its  new capabilities.

An additional sense in which numerical models can be tested is that observations
using other observatories have derived parameter values which fit adequately,
and {\it ASTRO-H} can test whether comparable fits are obtained with  these
same parameter values. This is new in the sense that  {\it ASTRO-H} is likely
to produce higher signal-to-noise and higher spectral resolution data than
previous observatories at energies greater than 3--5 keV. For example, the
{\it Chandra} HETG spectra of bright Seyfert galaxies such as NGC~3783 and
MCG\,$-$6-30-15 appear to be dominated by warm absorbers, and the spectra fit
adequately to several ionization parameter components
\citep{krongold03,holczer10}. These warm absorber components may be associated
features near the iron K lines, and these can be searched for using
{\it ASTRO-H}. 

More interesting is that {\it ASTRO-H} may allow testing of models which
include more realistic treatments of the dynamics and geometry of the outflow.
The list of such models includes those mentioned in the previous subsection,
plus the following:

1) Models for the time variability of the warm absorber, specifically  for one
case where a detection of correlated variability using {\it XMM-Newton} has
been  claimed: NGC~4051 \citep{krongold07}. Detection of variability depends on
a certain amount of luck, associated with catching a continuum variability
event.  Although there is no specific prediction for the variability in the 
iron band, it seems likely that more highly ionized material would  be
associated with more rapid variability than for the material  responsible for
the O, Ne, and Fe L absorption. Certainly, detection  of variability near the
iron K lines on time scales longer  than claimed from {\it XMM-Newton} in this object, or a
strong upper limit to variability on the {\it XMM-Newton} time scale, would represent a
strong test  of the model advanced for the warm absorber in NGC~4051 by
\citet{krongold07}.

2) Models which consider details of the 'geometry' of the  material responsible
for X-ray spectral features, i.e. those of \cite{sim10, sim12}. These models
make specific predictions for the iron K region, such  as the presence of
apparent UFO features.

3) It has been suggested that warm absorbers can mimic the apparent 
relativistically broadened iron lines in objects such as MCG\,$-$6-30-15
\citep{miller09}. If so, narrow features associated with highly ionized iron are
likely and should be detectable with {\it ASTRO-H}.

4) \cite{holczer07} suggest that the ionization parameter distribution of warm
absorber gas is determined by the effects of thermal instability. If so, there
should be additional material at ionization parameters greater than have been
previously detected, since this gas will be even more thermally stable than the
material detected  so far. Such material will be observable primarily in iron,
since it is  more resistant to ionization than other abundant elements.
{\it ASTRO-H} will  detect or constrain the existence of this gas, and so
indirectly test  the thermal instability hypothesis.

\subsection{Targets \& Feasibility}

AGN targets which will lend themselves to model testing are generically those
for which the data will have the  best signal-to-noise ratio in the energy band
most  favourable to {\it ASTRO-H}, i.e. the region of maximum SXS sensitivity
above 5 keV. These include the well known targets which are already listed as
likely targets and have been simulated elsewhere in this white paper, such as
MCG\,$-$6-30-15 and NGC~3783 (Sect.~\ref{sect:3783}) and NGC~5548
(Sect.~\ref{sect:5548}).

\section{Other types of sources with potential (narrow-band) outflow features} 

\subsection{Background and Previous Studies}

\subsubsection{BAL quasars}

Some quasars (QSOs) show fast outflowing gas features in their UV spectra. The
measured line blue-shifts well exceed 1000 km~s$^{-1}$ and can be as large as
$\sim 3-6\times 10^4$~km~s$^{-1}$ ($\sim$0.1--0.2c). Depending on the intrinsic
velocity width of the lines, these QSOs are classified as broad absorption line
(BAL; FWHM$>$2000 km~s$^{-1}$), mini-BAL (500 km~s$^{-1}$ $<$ FWHM $<$ 2000
km~s$^{-1}$) and narrow absorption line (NAL; FWHM$<$500 km~s$^{-1}$). In the
X-rays BAL QSOs are intrinsically weak and only for a few sources it has been
possible to perform a detailed spectral analysis \citep[e.g.,][]{chartas02}. On
the other hand, Mini-BAL QSO are a more promising class of sources as they
display a higher flux which translates in a higher quality X-ray spectrum. From
the few examples we know, these objects have complex spectra, characterized by
multicomponent warm absorbers, often with large column density and outflowing at
high velocity. Also ultra-fast-outflows have been reported for some of these
sources \citep{giustini11}.

\subsubsection{Partially covered sources}

Partially covered sources have been routinely detected and studied with current
instruments. The nature of the coverer is in general not known. Degeneracy
exists between a partial covering model and a reflection model, which can mimic
the same spectral shape, especially below 10 keV. A higher energy coverage is of
great help in trying to disentangle between the two scenarios. In some sources,
for example NGC~1365 \citep{risaliti09}, Mrk~766 \citep{miller07}, one or more
partial covering layers of gas have been claimed. The ionization structure of
these partial covering clouds is complex and sometimes they have been observed
through eclipses of the X-ray source. During the occultation from the cloud, a
cold absorber is often associated with a highly ionized gas (e.g., Fe~XXV and
Fe~XXVI), possibly suggesting a cometary structure. The study of the several
layers which (temporarily) obscure the source is important for understanding the
geometry of the gas around the black hole and how the gas layers are shielding
each other from the primary radiation. This can provide an important
observational benchmark for theoretical models of accretion disk winds (e.g.
launching and heating mechanisms).

\subsubsection{Compton thick outflows}

A related topic to partial covering sources and ultra-fast outflows is that of
Compton thick absorbers. In some remarkable sources, such as PDS~456, a Compton
thick absorber has been detected. This source is classified as a classical
Seyfert~1 galaxy (i.e. not a BAL QSO or a Seyfert~2). The absorber displays very
deep features from highly ionized iron possibly with multiple velocity
components. The features indicate that the gas is outflowing from the nucleus of
this galaxy at the relativistic speed of $\simeq$0.3~$c$. These outflows
possibly originate from the accretion disk itself and can have important
consequences in evaluating the feedback from quasars. Indeed, as shown in
Equation (\ref{eqn:mdot}), the kinetic luminosity carried by this plasma is a
strong function of the outflow velocity ($\propto \mathrm{v}^{3}$) and linearly
dependent on the column density ($N_{\rm H}\sim 10^{24}$\,cm$^{-2}$ in this
object).

\subsection{Prospects \& Strategy}

\begin{footnotesize}
\begin{table}[!htbp]
\begin{center}
\caption{\label{t:log} Possible targets for {\it ASTRO-H}}
\begin{tabular}{lccllll}
\hline
\hline
Name & type& redshift & EW (FeXXVI)& Ref. & Flux$_{(2-10 \rm keV)}$ &Ref.\\
&&& eV && erg\,cm$^{-2}$\,s$^{-1}$ &\\
\hline
PG\,1126-041 & miniBAL w UFO& 0.06 & 10.2$\pm$0.1 & $a$ & 3$\times10^{-11}$ & 2\\
\hline
Mrk\,766 & PC + HIG & 0.013 & 40--60 FeXXV & $b$ & 2$\times10^{-11}$ & 1\\ 
NGC\,1365 & PC +HIG & 0.005 & 100$\pm$60 FeXXVI\,K\,$\beta$ & $c$ & 1.5$\times10^{-12}$& 1\\
\hline
PDS\,456 & CT +UFO & 0.184 & 129$\pm$48 & $d$ & 3.5$\times10^{-12}$ &1\\
\hline
\end{tabular}\\
\end{center}
Notes:\\ 
PC: partial covering; BAL: broad absorption line quasar; UFO: ultra-fast
outflow; HIG: highly ionized gas; CT: Compton thick. Fluxes are 
in erg\,cm$^{-2}$\,s$^{-1}$\\
References for equivalent width (EW): 
$a$ \citet{giustini11}; 
$b$ \citet{risaliti11}; 
$c$ \citet{risaliti09}; 
$d$ \citet{reeves09}; 
References for flux measurements: 1: Tartarus data base; 2: 
\citet{giustini11}
\end{table}
\end{footnotesize}

The general prospects and strategy for these AGN are very similar to those
previously discussed in \S1.2 and \S2.2 for the ionized gas in ``normal'' and
``ultra-fast'' outflows, respectively. In particular, the large effective area
above 6~keV will allow to study a series of transitions for each component (e.g.
FeXXVI 1s--2p and 1s--3p and the K-edge). This will help in characterizing
optically thick plasmas where the main transitions are probably saturated.
Plasmas with large column densities do display features in the iron K$\alpha$
region even for low values of the ionizing parameter (e.g. lines from
Fe~XVIII-Fe~XXII). Therefore, this spectral region is then crucial for
disentangling the different gas components. The broad-band coverage and high
effective area of {\it ASTRO-H} will be very important to distinguish between
reflection and partial covering models (overlap with science case of reflection
in AGN). Finally, time-resolved or monitoring studies will be fundamental for
the determination of the covering fraction as a function of time and for
possible occultation events, such as X-ray eclipses.

\subsection{Targets \& Feasibility}

\subsubsection{BAL QSOs}
We present here a simulation for PG~1126-041. The required exposure time to
detect Fe~XXV at 5$\sigma$ is 400~ks. The input parameters for the simulation
come from \citet{giustini11}. These consist of a double powerlaw, one of which
is absorbed by two gas components. See Table~\ref{tab:pg1126} and
Figure~\ref{fig:pg1126}.

\begin{table}
\begin{center}
\caption{Parameters for PG~1126-041. Column densities are in units of cm$^{-2}$,
velocities in km\,s$^{-1}$ and fluxes in erg\,cm$^{-2}$\,s$^{-1}$.}
\label{tab:pg1126}
\begin{tabular}{|ll|}
\hline
$N_{\rm H}^1$ & $1.48\times10^{23}$\\
$\xi^1$ & 1.66\\
$N_{\rm H}^2$ & $7.5\times10^{23}$\\
$\xi^2$ & 3.4\\
$v_{\rm out}$ & -16200\\
$F_{\rm 0.2-10 keV}$ & $8.5\times10^{-11}$\\
\hline
Results:&\\
\hline
$N_{\rm FeXXV}$ & $1.23\pm0.24\times10^{19}$\\
$N_{\rm FeXXVI}$ & $2.7\pm1.1\times10^{18}$\\
\hline
\end{tabular}\\
\end{center}
\end{table}

\begin{figure}[!htbp]
\begin{center}
\resizebox{0.6\hsize}{!}{\includegraphics[]{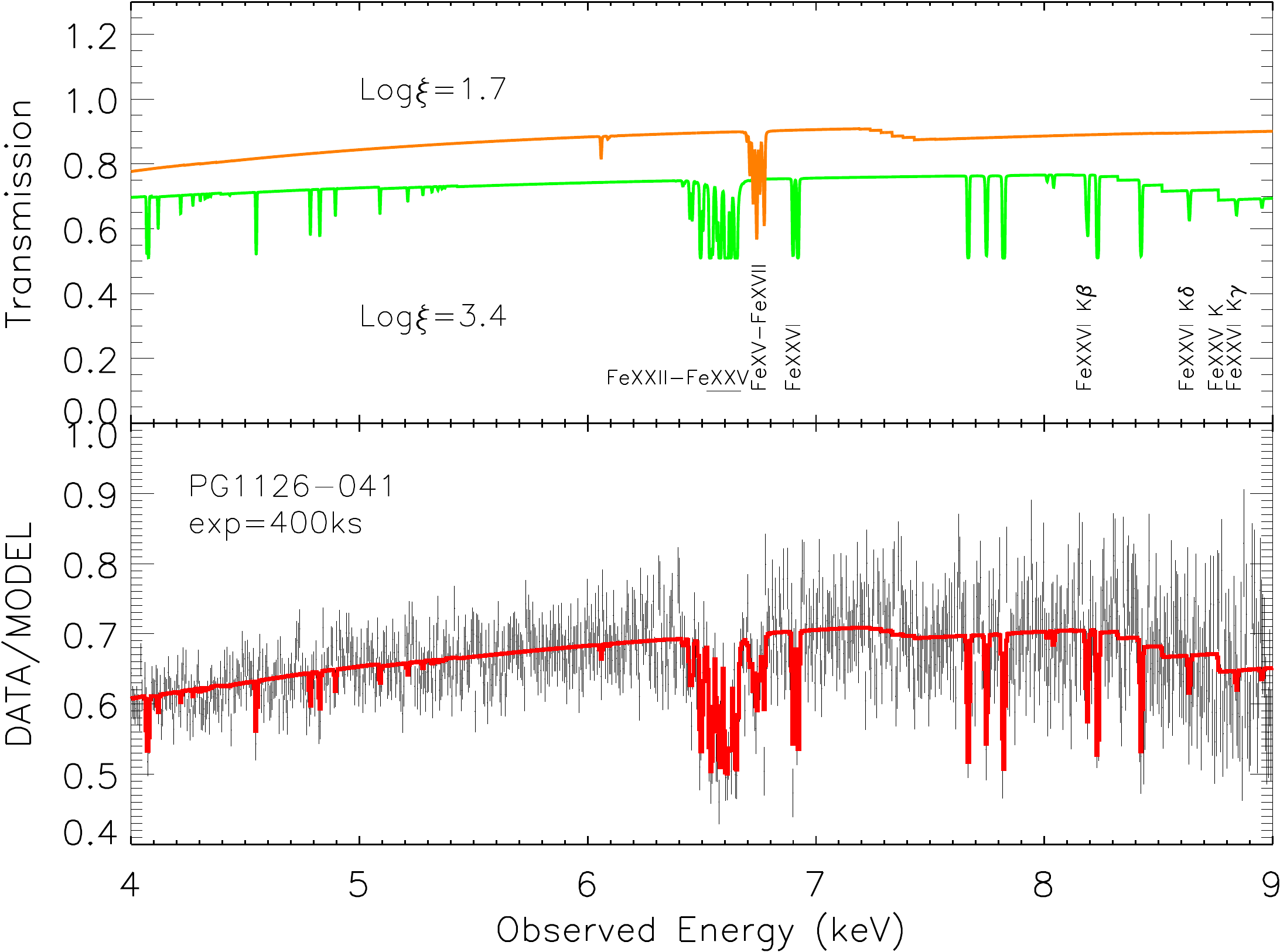}}
\end{center}
\caption{Simulated spectrum of PG~1126-041 for a 400~ks observation.}
\label{fig:pg1126}
\end{figure}



\subsubsection{Compton thick outflows}

PDS~456 ($z$ $=$ 0.184) has been well known as the most luminous AGN ($L_{\rm
bol}$ $\sim$ 10$^{47}$ erg s$^{-1}$) in the local universe. In spite of the fact
that more luminous AGNs tend to show less variability in the X-ray band, PDS~456
exhibits rapid and large amplitude flux variation accompanied by the spectral
changes on time scales of hours to years. Figure~\ref{fig:pds456_hist}, which is
referred from \citet{behar10}, shows the various spectra of PDS~456 observed by
several X-ray satellites. Each spectrum is divided by a power-law with
$\Gamma=2$ absorbed by Galactic column. 

\begin{figure}[!htbp]
\begin{center}
\resizebox{0.5\hsize}{!}{\includegraphics[]{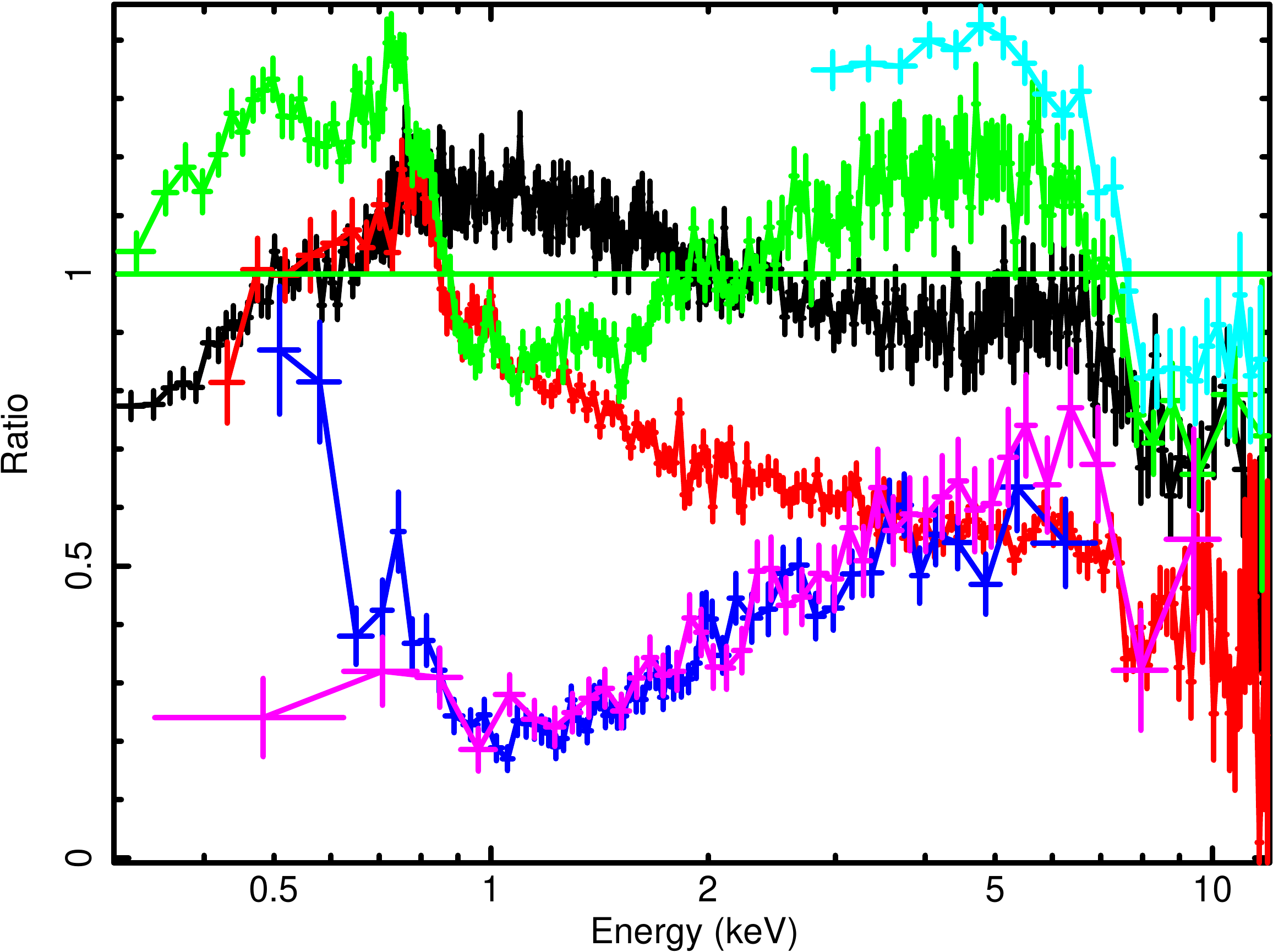}}
\end{center}
\caption{Various spectral state of PDS~456 \citep{behar10}. 
{\it XMM-Newton} observation in 2007 (black), {\it Suzaku} in 2007
  (red), {\it XMM-Newton} in 2001 (green), {\it Chandra} in 2003
  (blue), {\it ASCA} in 1998 (magenta), and {\it RXTE} in 2001 (cyan).}
\label{fig:pds456_hist}
\end{figure}

{\it ASCA} and {\it RXTE} observations revealed the presence of a strong
edge-like feature in the iron K-shell band \citep{reeves00}. \citet{reeves03}
reported the detection of blueshifted ($\sim$50\,000 km\,s$^{-1}$) iron K-shell
absorption features based on the spectrum obtained from the {\it XMM-Newton}
observation in 2001. Furthermore, {\it Suzaku} observation in 2007 also revealed
the significant absorption features near 9 keV in the source rest frame
\citep{reeves09}. If this absorption is due to blueshifted resonance transition
of hydrogen-like iron, then the outflow velocity reach nearly 30\% of the speed
of light. 

In this subsection, we present the results on the SXS simulation based on the
{\it Suzaku} spectrum obtained in 2007. An unprecedented energy resolution of
SXS in the iron band allows us to strictly determine the outflow velocity,
ionization states and the column density of the gas.

\begin{figure}[!htbp]
\begin{center}
\resizebox{0.6\hsize}{!}{\includegraphics[]{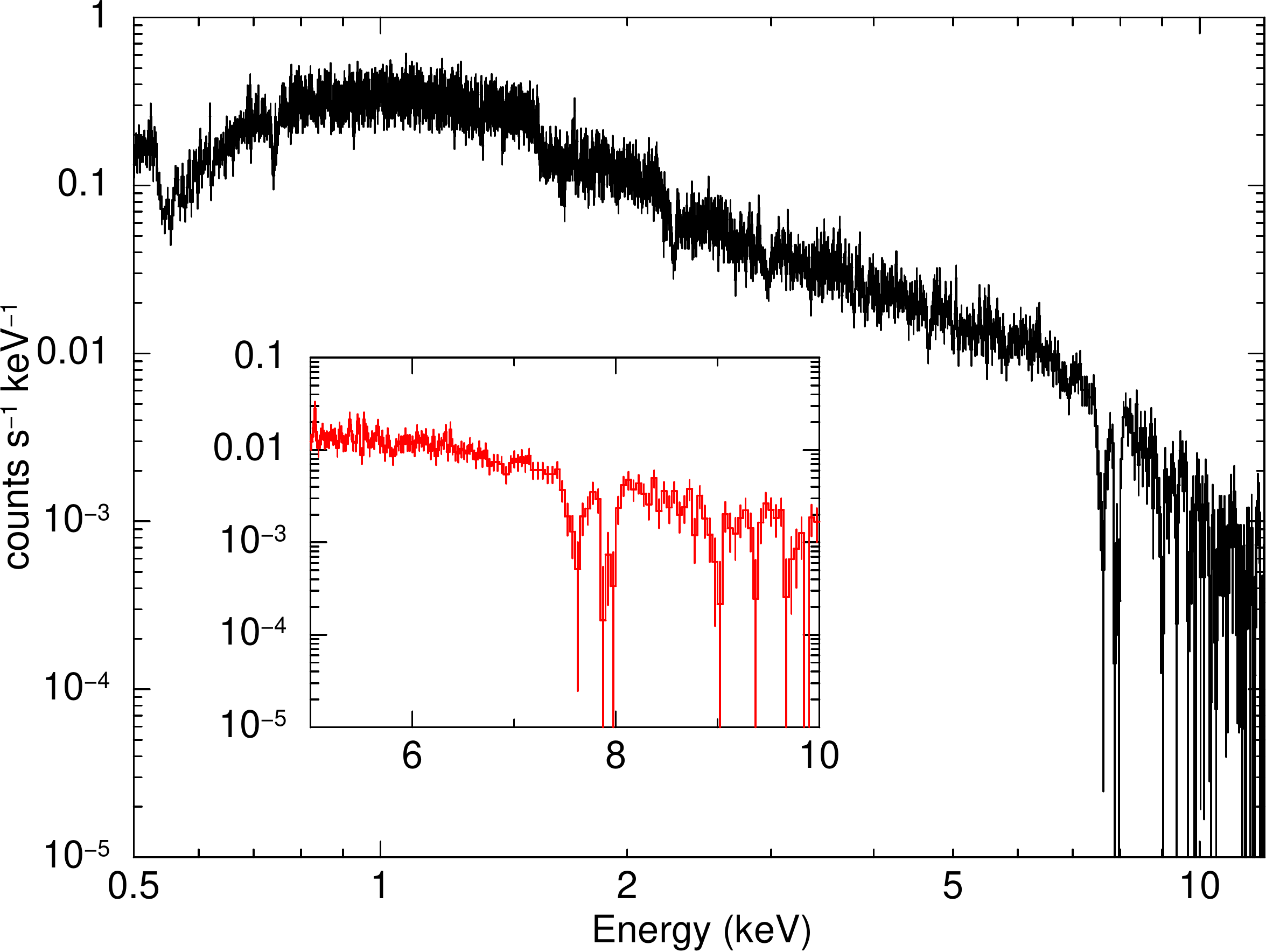}}
\end{center}
\vspace{-5mm}
\caption{Simulated spectrum of PDS~456 for 100~ks exposure. The inset shows an
enlargement of the iron line region.}
\label{fig:pds456_sim_spec}
\end{figure}

Figure~\ref{fig:pds456_sim_spec} shows the simulated spectrum with 100~ks
exposure. We used the redistribution matrix function (RMF) with 5 eV resolution,
which is the current best-estimated version
(ah\_sxs\_5ev\_basefilt\_20100712.rmf), and the ancillary response file (ARF)
for the point source (sxt-s\_120210\_ts02um\_intallpxl.arf). The assumed model
function is as follows: 
\begin{equation}
F(E) = \exp({-\sigma (E)N_{\rm H}^{\rm Gal}})\times({\rm WA}(\xi _{\rm WA},N_{\rm
  H}^{\rm WA},v_{\rm out})\times A_{\rm pl}E^{-\Gamma} + {\rm REF}(\xi _{\rm
  REF}, A_{\rm REF})),
\end{equation}

where WA and REF mean the absorption by warm absorber (XSTAR analytic model,
WARMABS in XSPEC) and the ionized reflection component \citep[REFLIONX in
XSPEC;][]{ross05,ross99}, respectively. $N_{\rm H}^{\rm Gal}$  and $N_{\rm
H}^{\rm WA}$ represent the column density of Galactic and warm absorber,
respectively. The ionization parameters of the absorber and the reflector are
expressed as $\xi _{\rm WA}$ and $\xi _{\rm REF}$, respectively. $v_{\rm out}$
is outflow velocity of the absorber, and the turbulent velocity of the absorber
was assumed to be 2\,500 km\,s$^{-1}$ according to Behar et al. (2010). $A_{\rm
pl}$ and $A_{\rm REF}$ are  normalization of power-law and reflection
components, respectively. The assumed parameters are summarized in
Table~\ref{tab:pds456_para}. 

\begin{footnotesize}
\begin{table}[!htbp]
\caption{\label{tab:pds456_para} Parameters in the simulated spectrum
  of PDS~456}
\begin{center}
\begin{tabular}{cccccccc}
\hline
\hline
$N_{\rm H}^{\rm Gal}$ & $\log\xi _{\rm WA}$ & $N_{\rm H}^{\rm WA}$ &
$v_{\rm out}$ & $A_{\rm pl}$ & $\Gamma$ & $\log\xi _{\rm REF}$ & $A_{\rm
  REF}$ \\
10$^{21}$ & erg\,s$^{-1}$\,cm & 10$^{24}$~cm$^{-2}$ & km\,s$^{-1}$ & & & erg\,s$^{-1}$cm & \\
\hline
2.1 & 3.5 & 1.0 & 90,000 & 3.1$\times$10$^{-3}$ & 2.25 & 3.0 & 3$\times$10$^{-9}$\\
\hline
\end{tabular}
\end{center}
\end{table}
\end{footnotesize}

\begin{figure}[!htbp]
\begin{center}
\resizebox{0.6\hsize}{!}{\includegraphics[angle=0]{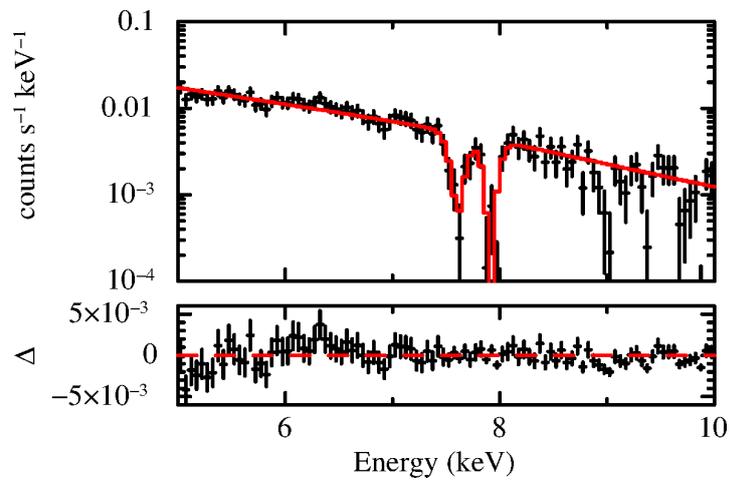}}
\end{center}
\vspace{-5mm}
\caption{5--10 keV spectrum of PDS~456 fitted with a single power-law and two
  negative Gaussians.}
\label{fig:pds456_fit}
\end{figure}

In order to estimate the significance of the absorption feature, we
fitted the 5--10 keV spectrum with a single power-law and two negative
Gaussian models. Figure~\ref{fig:pds456_fit} upper panel shows the
data (black) and model (red) we applied, and the residuals are
represented in the lower panel. Fitting results are summarized in
Table~\ref{tab:fit_result}. Both lines can be detected at highly
statistically significance level ($>$ 5$\sigma$). If these lines are
identified as H- and He-like iron, then the outflow velocity can be
determined to be 34.4$\pm$0.3\% of speed of light.

\begin{footnotesize}
\begin{table}[!htbp]
\caption{\label{tab:fit_result} Parameters in the simulated spectrum of PDS~456}
\begin{center}
\begin{tabular}{cccccccc}
\hline
\hline
(1) & (2) & (3) & (4) & (5) & (6) & (7) & (8) \\
$E_1$ & $\sigma _1$ & $F_1$ & $EW_1$ & $E_2$ & $\sigma _2$ & $F_2$ & $EW_2$ \\
\hline
9.37$\pm$0.02 &  90$^{+13}_{-7}$  & -4.9$^{+1.0}_{-0.4}$ & 0.25$^{+0.01}_{-0.02}$ & 
9.01$\pm$0.02 & 101$^{+28}_{-11}$ & -5.3$^{+0.8}_{-0.5}$ & 0.24$^{+0.03}_{-0.04}$  \\
\hline
\end{tabular}\\
\end{center}
{\small
Notes:\\ 
(1),(5); line center energy at the source rest frame in unit of keV. \\
(2),(6); line broadening in unit of eV.\\
(3),(7); line flux in unit of 10$^{-6}$ photons\,cm$^{-2}$\,s$^{-1}$.\\
(4),(8); line equivalent width in unit of keV. Errors on the
equivalent width are quoted at 68\% confidence. 
}
\end{table}
\end{footnotesize}

\clearpage
\begin{multicols}{2}
\end{multicols}
\end{document}